\documentstyle[12pt,aaspp4]{article}
\catcode`\@=11 
\def\lsim{\mathrel{\mathpalette\@versim<}}
\def\gsim{\mathrel{\mathpalette\@versim>}}
\def\be{\begin{equation}}
\def\ee{\end{equation}}

\def\rtr{r_{\rm tr}}
\def\mdot{\dot{m}}
\def\Mdot{\dot{M}}\def\mdotcr{\dot{m}_{\rm crit}}
\def\msun{M_{\odot}}
\def\@versim#1#2{\vcenter{\offinterlineskip
        \ialign{$\m@th#1\hfil##\hfil$\crcr#2\crcr\sim\crcr } }}
\catcode`\@=12 

\eqsecnum
\begin{document}
\date{}
\title{Spectral Transitions in Cyg X-1 and Other Black Hole X-Ray Binaries}
\author{Ann A. Esin, Ramesh Narayan}
\affil{Harvard-Smithsonian Center for Astrophysics, 60 Garden St, 
Cambridge, MA 02138}
\author{Wei Cui}
\affil{Center for Space Research, Massachusetts Institute of Technology, 
Cambridge, MA 02139}
\author{J. Eric Grove}
\affil{Naval Research Laboratory, Washington, DC 20375-5352}
\author{Shuang-Nan Zhang}
\affil{ES-84, NASA/Marshall Space Flight Center, Huntsville, AL 35812}

\begin{abstract}

We show that the model proposed by Esin, McClintock \& Narayan (1997)
for the low state, intermediate state and high state of the black hole
soft X-ray transient, Nova Muscae 1991, is consistent with the
spectral evolution of the black hole X-ray binary, Cyg~X-1, during the
hard-to-soft state transition observed in 1996.  We also apply the
model to the outbursts of two other black hole X-ray transients, GRO
J0422+32 and GRO J1719$-$24.

\end{abstract}

\keywords{accretion, accretion disks -- black hole physics -- X-rays: stars} 

\section{Introduction}
\label{intro}

Galactic X-ray sources that contain accreting black holes are among
the most interesting objects in high energy astrophysics.  An X-ray
source is classified as a black hole X-ray binary (BHXB) if the
observed mass function, or other less direct dynamical evidence,
places the mass of the compact object above $3 \msun$ (examples are
A0620$-$00, 4U 1543$-$47, V404 Cyg, Nova Muscae 1991, GRO J1655$-$40, GS
2000+251, Cyg X-1, GRO J0422+32, H1705$-$250 and LMC X-3), or if the
spectral and temporal characteristics of the source are similar to
those of the well established BHXBs (e.g.  GX 339$-$4, 1E 1740.7$-$2942,
GRS 1915+105, GRO J1719$-$24).  Recent reviews of this subject are
given by Tanaka \& Lewin (1995), Tanaka \& Shibazaki (1996), and Liang
(1997).

BHXBs are known to exhibit five distinct X-ray spectral states,
distinguished by the presence or absence of a soft blackbody component
at $\sim 1$ keV and the luminosity and spectral slope of emission at
harder energies. Systems in the {\em low/hard state} have hard
power-law spectra with a photon index $\Gamma \sim 1.4-1.9$, an
exponential cutoff around 100 keV, and no (or only weak) evidence of a
soft thermal component.  On the other hand, {\em high/soft state}
spectra are dominated by a blackbody component with a characteristic
temperature $\sim 1$ keV (Tanaka \& Shibazaki 1996).  At higher
energies, high state spectra show a power-law tail, softer than in the
low state ($\Gamma \sim 2.2-2.7$), but often extending above
500 keV without an obvious turnover. Of these two spectral states, the
high state is generally more luminous, with $L_X \sim 0.2-0.3 L_{\rm
Edd}$, while systems in the low state have X-ray luminosities below
$\sim 0.1 L_{\rm Edd}$ (e.g. Nowak 1995); this trend is of course
based on systems for which distance and black hole mass estimates are
available.  Recently, {\em intermediate state} spectra have been
seen in several systems (e.g.  Nova Muscae, Ebisawa et al. 1994; GX
399$-$4, Mendez \& van der Klis 1997; Cyg X-1, Belloni et al. 1996),
which, as the name suggests, are intermediate both in spectral shape
and luminosity between the high and low state spectra.

Two other spectral states have been seen in some BHXBs.  At least two
sources, Nova Muscae and GX 339$-$4 (Ebisawa et al. 1994; Miyamoto et
al 1991), have been observed in the {\em very high state}, with
luminosities close to the Eddington limit.  This state has similar
spectral properties as the high state, but is characterized by higher
bolometric luminosity and more prominent hard tail, stronger
variability and the presence of $3-10$ Hz QPOs (e.g. van der Klis
1994; Gilfanov et al. 1993).  In addition, all transient BHXBs
(e.g. Nova Muscae, A0620$-$00, V404 Cyg) spend most of their time in a
{\em quiescent/off state}, characterized by a very low X-ray flux
(e.g.  McClintock, Horne \& Remillard 1995; Narayan, Barret \&
McClintock 1997a; Narayan, Garcia \& McClintock 1997b).

In the past several decades X-ray and $\gamma$-ray observations have
considerably improved our understanding of BHXBs, and
significant progress has been made in theoretical modeling of
different spectral states.  It has been generally accepted that the
soft thermal component originates in an optically thick,
geometrically thin accretion disk (Shakura \& Sunyaev 1973; Novikov \&
Thorne 1973), while the power-law high energy emission is produced by
Comptonization of soft photons in a hot corona (e.g.  Shapiro,
Lightman, \& Eardley 1976; Haardt et al. 1993; Melia \& Misra 1993).
However, the spatial relationship between the hot and cold medium in
the accretion flow has until recently been unclear.

Narayan, McClintock \& Yi (1996) and Narayan et al.(1997a) showed that
to produce the observed optical and X-ray emission in quiescent soft
X-ray transients (SXTs), the accreting gas must be primarily in the
form of an optically thin hot flow which extends out to a large
radius, of order several thousand Schwarzschild radii; beyond this
radius, the accretion occurs via a thin disk. These authors modeled the
hot accreting gas as an optically thin advection-dominated accretion
flow (ADAF).  This is an accretion solution which was discovered by
Ichimaru (1977) and Rees et al. (1982), and then rediscovered and
further studied by Narayan \& Yi (1994, 1995a, 1995b), Abramowicz et
al. (1995), Chen (1995), Chen et al. (1995) and others.  Hot optically
thin ADAFs exist only below a certain critical accretion rate,
$\mdotcr \sim 0.1$ in Eddington scaled-units.  Because the density of
the gas is low, the flow is unable to cool efficiently.  Instead, most
of the viscously dissipated energy is stored in the gas as thermal
energy of the gas particles and is advected into the black hole
through the event horizon.  As a result, the radiative efficiency of
an ADAF around a black hole is significantly lower than that of a
standard thin disk.  This makes the ADAF model especially appropriate
for modeling under-luminous quiescent systems.

The model proposed by Narayan et al. (1996, 1997a) for quiescent SXTs
was used by Hameury et al. (1997) to explain the light curve of the
transient source, GRO J1655-45, during its recent outburst.
Previously, Mineshige \& Wheeler (1989) and Huang \& Wheeler (1989)
had suggested that outbursts in transients are due to a thermal
instability in the thin accretion disk and extended a model developed
previously for dwarf novae to black hole transients.  However, if the
thin disk extends down to the last stable orbit in quiescence, the
predicted recurrence time between outbursts is inconsistent with
observations (Lasota, Narayan, \& Yi 1996; Mineshige 1996).  The
predicted and observed time scales are in better agreement if the thin
disk is truncated at a large transition radius, $R_{\rm tr}$.

In ADAF models of quiescent SXTs, the mass accretion rate is fairly
low, $\mdot \lsim 10^{-3}$ in Eddington units (Narayan et al. 1996,
1997a).  Narayan (1996) proposed that at higher accretion rates the
same model might naturally explain the more luminous spectral states
of black hole binaries.  This suggestion was worked out in detail by
Esin, McClintock \& Narayan (1997, hereafter EMN).  They showed that
the low state is similar to the quiescent state in its flow
configuration, but with a mass accretion rate roughly two orders of
magnitude higher, namely $0.05 \lsim \mdot \lsim 0.1$, where $\mdot$
is in Eddington units. Similar models were proposed for the low state
of Cyg X-1 by Shapiro et al. (1976) and Ichimaru (1977).  In
particular, Ichimaru (1977) was the first to recognize the importance
of advection dominated accretion for understanding BHXBs; sadly, his
paper was forgotten for many years.  At yet higher accretion rates, an
ADAF cannot be maintained at large radii; the transition radius
(i.e. the inner edge of the thin disk) then comes closer to the black
hole.  The spectra produced by such models resemble well the
intermediate state spectra observed in some systems.  Finally, for yet
higher mass accretion rates, the thin disk extends all the way down to
the last stable orbit and the ADAF is restricted to a corona above the
disk. Such a configuration describes the high state.

EMN demonstrated that their model, consisting of an inner ADAF
surrounded by an outer thin disk, can reproduce the low, intermediate
and high state spectra of BHXBs.  Moreover, they showed that the
existence of the critical accretion rate $\mdotcr$ above which the
ADAF solution disappears provides a natural explanation for spectral
transitions between the high and low state.  EMN modeled the outburst
of Nova Muscae 1991 using this scenario and showed that the model
successfully reproduces the observations of the system in the three
spectral states.

Recent observations of several BHXBs have confirmed the basic
conclusions of EMN.  By studying the fluorescent Fe K$\alpha$ line and
Compton reflection component in the low state spectrum of Cyg X-1,
Gierli\'nski et al. (1997a) concluded that the cold reflecting medium
does not extend down to the marginally stable orbit (for a
non-rotating black hole) in the low state.  The same conclusion was
also reached by Dove et al. (1997a, 1997b), who found that the
reprocessing of the hot radiation in the cold medium is too weak, and,
therefore, that the cold disk must be truncated relatively far outside
$3 R_{\rm Schw}$. A more direct confirmation came from Zhang et
al. (1997a, 1997b) and $\dot{\rm Z}$ycki et al. (1997) who studied the
transition between the low and high spectral states in Cyg X-1 and
Nova Muscae respectively, and found a larger value for the transition
radius in the low state than in the high state.  This confirms the
overall qualitative features of the EMN model.

We begin this paper in \S2 with a review of the predictions of the EMN
model; in particular, we discuss the spectral signatures predicted for
various state transition, and show how they can be used as diagnostics
when the available spectral information is incomplete.  In \S3 we
concentrate on the recent hard--soft spectral state transition in Cyg
X-1.  We show that the data are consistent with the basic scenario
described above, namely that during the transition Cyg X-1 went from
the low state, with a relatively large inner radius for the thin disk
and an accretion rate near the critical value, to a series of
intermediate states with smaller transition radii, and culminated in
the high state with $R_{\rm tr}= 3 R_{\rm Schw}$.  We show that this
model convincingly explains the overall spectral evolution of the
system, as well as reproduces its hard and soft state spectra fairly
well.

In \S4 we discuss the application of the model to another black hole
system, GRO J0422+32, during its 1992 outburst.  We demonstrate that the
spectrum of GRO J0422+32 near the peak of the outburst is well modeled
by a low state spectrum with $\mdot \sim \mdotcr$.  We then argue that
since the spectrum remained a hard power-law throughout
the outburst, the accretion rate never increased significantly above its
critical value.  Thus GRO J0422+32 never went into an intermediate or
high spectral state.  We show that this conclusion is consistent with
the observed anti-correlation between the cutoff energy and the X-ray
luminosity during the decay after the outburst. 

Finally, in \S5 we show that the EMN model also provides some clues to
the secondary outbursts observed in GRO J1719$-$24 (also known as
GRS~1716-249 or Nova Ophiuchi 1993).  Though the spectral evolution of
the system since its discovery in September 1993 is not
well-documented in the literature, we show that the known facts are
consistent with an outburst from the quiescent to the very high
spectral state, like those seen in ``typical'' SXTs (e.g. Nova Muscae or
A0620$-$00), followed by a series of low-high state transitions, similar
to those observed in Cyg X-1.  We conclude in \S6 with a summary.
 
\section{Spectral States and State Transitions of BHXBs}
\label{mainmodel}

We begin with a brief description of the model geometry and the
calculations (see EMN for details).

\subsection{ADAF Model Overview}
\label{model}

\subsubsection{Flow Geometry}
\label{geometry}
We consider a Schwarzschild black hole of mass $M = m \msun$ accreting
matter from its binary companion at a rate $\Mdot = \mdot
\Mdot_{Edd}$, where $\Mdot_{Edd} = L_{Edd}/(0.1 c^2) = 1.39\times
10^{18} m$ is the Eddington accretion rate computed for $10\%$
radiative efficiency.  The binary orbital angular momentum vector is
inclined at an angle $i$ to the observer's line of sight.

The basic picture of mass accretion via an ADAF in the context of
BHXBs was worked out in a series of papers by Narayan and
collaborators (Narayan et al. 1996; Narayan 1996; Narayan et
al. 1997a; Hameury et al. 1997; EMN).  In this scenario, the accretion
flow is divided into two distinct zones.  The inner part is modeled as
a hot optically thin ADAF extending down to the black hole horizon,
while the outer part consists of an optically thick, geometrically
thin disk with a hot corona (modeled as an ADAF) above it.  The
transition radius between the two zones, $\rtr = R_{\rm tr}/R_{\rm
Schw}$ (hereafter all radii are in Schwarzschild units), is one of the
model parameters.  Another less important parameter is the radius of
the outer edge of the accretion flow, $r_{\rm out}$.
 
Though the total mass accretion rate in the disk and corona is constant
at each radius (this is not strictly true in time-dependent flows, but
we make this assumption for simplicity), it is not possible to determine
from first principles the fraction of mass accreted through the corona.
This is because the coupling between the hot corona and the cold disk
through evaporation and thermal conduction (see e.g. Meyer \&
Meyer-Hofmeister 1994), is poorly understood.  We simply assume that
the coronal mass accretion rate is inversely proportional to the radius,
i.e.  
\be 
\mdot_c(r) = \mdot \left(\frac{\rtr}{r}\right),\ \ \ \ r > \rtr.  
\ee 
(The results are not very different if we assume other radial
profiles for $\mdot_c$.)

\subsubsection{Flow Dynamics and Energy Balance}
\label{dynamics}

Once the total accretion rate $\mdot$ and accretion rate in the corona
$\mdot_c(r)$ are set, all the relevant properties of the hot flow can be
computed. The ADAF is treated as a collection of spherical shells,
truncated near the pole to mimic the flattening of the density profile.
Each shell is characterized by its radial and azimuthal
velocity, gas density, electron and proton temperatures, magnetic field
strength, etc.  These quantities are determined by solving the global
dynamical conservation laws of mass, angular momentum, radial momentum and
energy, together with the radiative transfer problem. The
manner in which the solutions are obtained is described in detail in
Narayan, Kato \& Honma 1997c, Narayan et al. (1997a) and
EMN; here we present a brief summary and introduce the model
parameters. 

In each shell the viscous energy dissipation is computed using the
standard $\alpha$-prescription.  We assume that most of the energy is
deposited in the protons (see Bisnovatyi-Kogan \& Lovelace 1997,
Quataert 1997, Gruzinov 1997 and Blackman 1997 for recent discussions
of this assumption); the protons then reach nearly virial temperature
through viscous heating and adiabatic compression.  We further assume
that the electrons and protons are coupled only through Coulomb
collisions, which is inefficient in a hot, low-density plasma.  Thus,
only a fraction $(1-f)$ of the viscous energy is transferred to the
electrons, and the rest is carried inside the black hole horizon as
entropy of the gas.  The quantity $f$ is called the advection
parameter, and is solved for self-consistently.

The electrons in the flow cool via three processes: bremsstrahlung,
synchrotron radiation, and inverse Compton scattering.  To compute the
synchrotron emissivity of the gas (Mahadevan, Narayan \& Yi 1996), we
assume that the tangled magnetic field is roughly in equipartition
with the hot gas, so that the ratio of the gas to total pressure,
$\beta$, is of order 0.5.  In calculating the rate of cooling due to
Compton scattering we take into account the coupling between different
shells of the ADAF, as well as the interaction between the hot flow
and the thin disk (Narayan et al. 1997a, EMN).

The radial profiles of electron and proton temperature in the ADAF are
determined by demanding that the energy equations for both protons and
electrons are satisfied in each radial shell.  The rate of change of
entropy of the protons is set equal to the viscous energy dissipation
minus losses through Coulomb collisions with electrons.  For the
electrons, the entropy change is set equal to the energy gain through
Coulomb coupling with the protons, plus the fraction $\delta$ of the
viscous energy dissipation that goes directly into heating the
electrons, minus radiative cooling.  In our calculations we adopt
$\delta = 10^{-3}$, though the results are not sensitive to the exact
value of this parameter as long as $\delta \lsim 10^{-2}$. 

\subsubsection{Emission from the Thin Disk}
\label{thindisk}

The emission from the thin disk is calculated using the standard
multicolor blackbody method (e.g. Frank et al. 1992) corrected for
graybody effects due to electron scattering (e.g. Shimura \& Takahara
1995); in addition, we include the Compton reflection component due to
the scattering of hot ADAF photons incident on the disk surface, and
fluorescent iron K$\alpha$ line emission.  In EMN the thin disk was
treated as an infinitely thin plane.  Here we take into account the
finite disk thickness, which we assume to be equal to the local
pressure scale height.  The scale height itself is calculated
self-consistently including the effect of irradiation of the disk by
radiation from the ADAF.

The modified blackbody spectrum of the disk is computed using the
standard expression:
\be
L_{\nu,{\rm disk}}(r) dr = \left(\frac{1}{f_{\rm mbb}}\right)^4 
B_{\nu}(f_{\rm mbb}\ T_{\rm eff}(r)) \cdot 2 \pi \cdot 2 \pi r dr,
\ee
where $B_{\nu}$ is the Planck function, $T_{\rm eff}(r)$ is the
effective surface temperature of the disk at radius $r$ (which includes
both the local viscous dissipation as well as the fraction of the
irradiating flux which is absorbed), and $f_{\rm mbb}$ is a factor which
allows for modified blackbody effects; we set $f_{\rm mbb}$ equal to 1.7
following Shimura \& Takahara (1995). 

Not all photons incident on the thin disk are absorbed and
reprocessed.  A significant fraction of the incident energy is Compton
scattered in the upper disk layers and is effectively reflected back
with lower energy than that of the incident photons.  The result is a
broad peak centered at around 30 keV.  We compute this part of the
spectrum by convolving the incident spectrum with the angle-averaged
Greens function for Compton reflection of monoenergetic photons from a
cold, neutral, optically thick medium (White, Lightman \& Zdziarski
1988; Lightman \& White 1988).  The metal abundances are taken from
Morrison \& McCammon (1983).

Finally, we compute the strength and shape of the fluorescent iron
line produced through photoelectric absorption of the incident X-ray
photons with energies above the K-shell absorption edge, $E_K=7.1\
{\rm keV}$.  We compute the line strength using the empirical formula
given by George \& Fabian (1991), and the line profile taking into
account gravitational redshift and Doppler shifts due to rotation
(Fabian et al. 1989).  However, we ignore the effects of ray bending
in the vicinity of the black hole; this effect becomes important only
when the transition is below $r = 10$ (Fabian et al. 1989).

\subsection{Spectral States of BHXBs}
\label{states}

EMN calculated the spectra of the above model as a function of the two
main parameters, $\mdot$ and $\rtr$, and identified different regions of
parameter space with the known spectral states of black hole systems.
Here we summarize their scenario for the quiescent, low, intermediate
and high spectral states.  We ignore the very high state, which is not
relevant for this paper (this state was in any case, not explained very
convincingly by EMN). 

At low mass accretion rates ($\mdot \lsim 10^{-2}$) and relatively
large transition radii ($\rtr \gsim 10^3$) the model spectra strongly
resemble the quiescent states of transient black hole systems (e.g.
A0620-00, V404 Cyg, GRO J1655-40; see Narayan et al. 1996,
Narayan et al. 1997a; Hameury et al. 1997).  In this region of
parameter space, the optical/UV part of the spectrum is dominated by
self-absorbed synchrotron emission from the ADAF (NBM) and the X-rays
are produced by bremsstrahlung emission and Comptonized synchrotron
photons.  The contribution from the thin disk is negligible.  The
spectral slope in the X-ray band varies strongly with the accretion
rate (see Figure 1a).  For $\mdot \lsim 10^{-3}$, bremsstrahlung
dominates and the spectral slope is roughly constant with photon index
$\Gamma \sim 1.7$ (N.B. in EMN the photon index was denoted by the
symbol $\alpha_N$).  In the range $10^{-3} \lsim \mdot \lsim 10^{-2}$,
the X-ray spectrum is dominated by Compton scattered synchrotron
photons, and the photon index increases to $\Gamma \sim 2.2$,
producing a steeper spectrum. The radiative efficiency of the ADAF is
proportional to $\mdot$ (Narayan \& Yi 1995b), so that the quiescent
state spectra have very low luminosities, $L \sim L_{\rm Edd}
(\mdot^2/0.1)$, consistent with what is observed.

Models with $\mdot \gsim 10^{-2}$ and $\rtr > 100$ are associated with
the low state of BHXBs.  In this regime, the X-ray part of the spectrum
is formed by Comptonized synchrotron photons.  Therefore, with
increasing $\mdot$ as the optical depth and Compton $y$-parameter
increase the spectral slope flattens (Figure 1a).  The emission is
dominated by high energy photons at $\sim 100$ keV.  The thin disk may
contribute significantly to the observed emission in the optical/UV
band, though the exact contribution depends on the value of $\rtr$
(compare Figures 1a and 1b).  However, the shape of the X-ray spectrum
is not sensitive to the location of the transition radius, as long as
$\rtr \gsim 100$.  In fact, even with $\rtr \sim 30$, the slope in the
1-10 keV range remains hard, and the only noticeable difference is in
the appearance of an ultrasoft disk component below 1 keV, the presence
of a reflection bump at $\sim 30$ keV and the strength of iron
fluorescent line. 

It has been demonstrated in several papers that two-temperature ADAFs
exist only for accretion rates below some critical value $\mdotcr$
(Ichimaru 1977, Rees et al. 1982, Narayan \& Yi 1995b, Chen et al.
1995, EMN).  Above $\mdotcr$ the accreting gas cools so rapidly, that
the flow cools down to form the standard optically-thick Shakura-Sunyaev
disk.  EMN demonstrated that $\mdotcr$ is a function of the transition
radius between the hot and cold flow, increasing slightly with
decreasing $\rtr$.  Thus, they proposed that when the accretion rate in
BHXBs increases above the critical value for a given value of $\rtr$,
the outer parts of the hot flow start cooling down to form a thin disk,
i.e. $\rtr$ decreases in an attempt to keep $\mdot$ at $\mdotcr$.  When
$\rtr$ falls below about 30, the model spectra resemble the intermediate
state spectra observed in several black hole systems, e.g. Nova Muscae
(Ebisawa et al. 1994), Cyg X-1 (Belloni et al. 1996), GX 339$-$4 (Mendez
\& van der Klis 1997). 

Since $\mdotcr$ changes by at most $\sim 15\%$ while $\rtr$ decreases by
a large factor the bolometric luminosity is roughly the same for all
intermediate state spectra.  The spectral shape, however, is strongly
dependent on the value of $\rtr$.  For $\rtr > 30$ the X-ray spectra are
similar to those in the low state (large $\rtr$), with $\Gamma \sim
1.5$, while by the time $\rtr \sim 3$, the spectrum is dominated by the
soft blackbody component from the disk at $\sim 1$ keV, with a steep
power-law tail characterized by $\Gamma \sim 2.3$. 

When the thin disk extends down to the last stable orbit, the ADAF is
restricted to the corona above the disk.  EMN associate this flow
configuration with the high state, in which blackbody radiation from
the disk dominates the spectrum and the corona produces only a weak
high energy tail.  Because the electron temperature in the ADAF is
rather low in the high state configuration, the power-law tail in our
model spectra turns over at $\sim 100$ keV, while the observed spectra
extend to $\sim 500$ keV without a cut-off (e.g. Grove et al. 1997).
This discrepancy either indicates the presence of non-thermal
electrons, or more likely is due to bulk motion Comptonization of the
thin disk photons, as suggested by Ebisawa, Titarchuk \& Chakrabarti
(1996).  The radial velocity of the ADAF gas is on the order of the
free fall velocity, and since the electron temperature in the high
state decreases, the bulk motion Comptonization (which we do not take
into account in our model) might become more important than thermal
Comptonization.

In their model for the spectral evolution of Nova Muscae during its 1991
outburst, EMN assumed that the plateau in the lightcurve of the
system observed between days 130 and 200, corresponds to the transition
from the high to the low spectral state, with $\rtr$ increasing from 3
to $\sim 10^4$ (the upper value was determined from the quiescent state
spectrum, see Narayan et al. 1996).  This assumption was made for the
sake of simplicity, since the data used in the paper were not good
enough to place better constraints on $\rtr$ in the low state.  Recent
detailed analysis by $\dot{\rm Z}$ycki et al. (1997), seems to imply
that by day 200, $\rtr$ increased only up to $\sim 20$.  To reconcile
this result with the value of $\rtr$ in quiescence, we must conclude
that the thin disk continued to evaporate after day 200, while the mass
accretion rate dropped from its value in the low state, to that in
quiescence.  As pointed out above, as long as $\rtr \gsim 30$ in the low
state, the exact position of the transition radius does not
significantly change the X-ray spectrum (compare Figures 1a and 1b).
Thus this slight modification of the model has no effect on the main
results of EMN. 

In the discussion above we have outlined how the spectrum of the ADAF
plus the disk changes as a function of $\mdot$ and $\rtr$.  The other
parameters, such as $m$, $r_{\rm out}$, $i$, $\alpha$ and $\beta$, do
not significantly affect the main features of the model spectra.  For
example, changing the mass of the accreting black hole affects only
the normalization of the spectrum and the effective temperature of the
thin disk ($T_{\rm eff}\propto m^{-1/4}$, Frank et al. 1992).  Larger
values of $\alpha$ and $\beta$ produce slightly harder low state
spectra, and raise the value of $\mdotcr$ (see Narayan 1996;
Narayan et al. 1997a and EMN for detailed discussion), but the overall
behavior of the model is unaffected.

In the calculations presented below, we generally set $\beta=0.5$ and
$\alpha=0.3$, which we consider to be the most natural values for
these parameters.  The former simply restates our assumption of
equipartition between magnetic and gas pressure, a common assumption
in many areas of high-energy astrophysics.  For the viscosity
parameter $\alpha$ we follow the prescription suggested by Hawley \&
Balbus (1996), $\alpha \sim c (1-\beta)$ with $c$ between 0.5 and 0.6.
For the other two parameters, $m$ and $i$, we adopt values suggested by
observations of the systems we model.

\subsection{Spectral Signatures During State Transitions}
\label{transition}

Though some black hole candidates are always observed in the same
spectral state (e.g. LMC X-3 and 1E 1740.7$-$2942), the majority of
observed BHXBs exhibit some degree of long-term spectral variability.
The most dramatic behavior is observed in soft X-ray transients (also
known as ``X-ray novae''), which in some cases have been detected in
all five spectral states, including the ``very high state'' not
discussed in this paper.  A typical example of such a system is Nova
Muscae (e.g. Ebisawa et al.  1994, EMN).  Most other black hole
systems exhibit two or three spectral states.  For example, Cyg X-1
routinely switches between the low state and high state, passing
through a series of intermediate states (e.g. Phlips et al. 1996;
Zhang et al. 1997a, see also review by Tanaka \& Lewin 1995 and
references therein); the high and low states were also detected during
OSSE observations of GRO J1719$-$24 (Grove et al. 1997).  On the other
hand, the transient source GRO J0422+32 has been seen only in the
quiescent state and low state (Grove et al. 1997; Tanaka \& Shibazaki
1996 and references therein).  However, regardless of the number of
distinct states observed, we believe most systems are described by the
model outlined in \S2.2.

Figures 1a and 1b illustrate the model predictions for the various
spectral states.  Shown are a sequence of spectra computed for $m=9$,
$i=40^{\circ}$ and values of $\rtr$ and $\mdot$ as stated in the figure
caption.  Starting with the lowest spectrum, which corresponds to the
quiescent state, one can follow the spectral evolution all the way to
the luminous high state.  Apart from the monotonic increase in the
overall luminosity, the most striking changes are seen in the spectral
slope in the X-ray band.  The spectrum is hard in the quiescent state,
becomes softer as the mass accretion rate increases (the photon index
reaches its maximum $\Gamma \sim 2.2$ for $\mdot \sim 10^{-2}$), and
switches again to a very hard spectrum in the low state.  During the
transition from the low to the high state, the spectrum softens again
and continues to soften with increasing $\mdot$ in the high state.  Note
that the exact value of $\rtr$ in the low and quiescent states changes
only the characteristic frequency of the thin disk component (generally
located in the UV or optical part of the spectrum) and not the shape of
the X-ray emission. 

A less obvious, but nonetheless important change takes place in the
position of the high energy thermal cut-off in the spectrum.  In the
quiescent state, the cooling of the flow is very inefficient and the
electron temperature in the inner regions of the ADAF is relatively
high, reaching $4-6 \times 10^{9}$ K.  However, with increasing
$\mdot$ the electron temperature drops due to increased cooling, and
the exponential cut-off occurs at lower energies.  The resulting
relationship between the average electron temperature of the flow
inside $10 R_{\rm Schw}$, and the X-ray luminosity at 100 keV (the
band which is most sensitive to changes in $\mdot$), is shown in
Figure~2.  Clearly, during the transition from the quiescent to the
low state, $T_e$ and the hard X-ray luminosity are {\em anti-correlated}.

During the transition through the intermediate states, when $\rtr$
decreases without significant change in $\mdot$, $T_e$ continues to
fall.  As the blackbody disk component becomes more prominent, the hot
accretion flow is cooled progressively more efficiently by the soft
photons from the thin disk.  Both the hard X-ray flux and the
temperature decrease. Therefore, in the intermediate spectral state, 
the flux at 100 keV and $T_e$ are {\em correlated} with each other, as
clearly shown in Figure 2.

Though the spectra corresponding to different states are very distinct
when compared over a wide range of frequencies, say from optical to
X-rays, in real life one generally has much more limited information,
e.g. just the hard X-ray flux. When the data are restricted to a
relatively narrow energy band, different spectral transitions may look
rather similar.  The relationship between $T_e$ and the hard X-ray flux
illustrated in Figure 2 could then be used as a diagnostic to determine the
nature of the transition.

\section{Application to Cyg X-1 1996 State Transition}
\label{cygx1}

Cyg X-1, discovered by Bowyer et al. (1965), is one of the oldest known
X-ray sources and perhaps the most thoroughly studied Galactic black
hole candidate.  The system consists of a compact object orbiting around
an O9.7~Iab supergiant (Walborn 1973; Gies \& Bolton 1986a) with a
period of 5.6 days.  The X-ray emission is most likely powered by
focused wind accretion from the optical companion (Gies \& Bolton
1986b).  The extinction towards Cyg X-1 places this system at a distance
of $\gsim 2.5$ kpc (Margon \& Bowyer 1973; Bregman et al. 1973); we use
2.5 kpc for comparisons between the model and observations. 

From the orbital motion of the primary, the mass function of the
compact object is measured to be $f(M) = 0.25\pm 0.01 \msun$ (Gies \&
Bolton 1982), and recent estimates (e.g. Herrero et al. 1995) place
the mass of the optical companion at $18 \msun$.  Unfortunately, the
orbital inclination of the system is not well known; the values quoted
in the literature vary between $25^{\circ}$ and $67^{\circ}$ (e.g.
Dolan 1992; Ninkov et al. 1987 and references therein).  This
uncertainty implies that the mass of the black hole candidate can be
anywhere in the range between $15\msun$ and $5.5\msun$.  For our
calculation we adopt $m=9$ and $i=40^{\circ}$.

\subsection{Low/Hard Spectral State}
\label{cygx1_low}
Cyg X-1 spends most of its time in the hard state, which is
characterized by a hard power-law X-ray spectrum with a photon index
$\Gamma\sim 1.4$.  The power-law is modified by an exponential cutoff
with e-folding energy $E_f \sim 150$ keV (e.g.  Phlips et al. 1996;
Dove et al. 1997b; Gierli\'nski et al. 1997a, and references therein),
and perhaps some soft blackbody emission with characteristic
temperature $\sim 0.1-0.2$ keV (e.g. Balucinska-Church et al. 1995;
Ebisawa et al. 1996).  In some hard state spectra of Cyg X-1 there is
also a strong indication of spectral hardening above 10 keV, usually
interpreted as a Compton-reflection component from a cold disk (Done
et al. 1992; Gierli\'nski et al. 1997a).  Another evidence for the
presence of the disk is the detection of an Fe K-edge at $\gsim 7$ keV
as well as a narrow Fe K$\alpha$ fluorescence line (e.g.  Barr, White
\& Page 1985; Ebisawa et al. 1996; Gierli\'nski et al. 1997a).

Historically, the combination of hard power-law continuum and
reflection features has been interpreted in terms of a cold disk with
a slab-like hot corona model (e.g. Haardt et al. 1993 and references
therein; though see also Shapiro et al. 1976 and Ichimaru 1977), where
the hard X-ray emission is produced by inverse Compton scattering of
the disk photons in the corona, and the irradiation of the thin disk
by the hot photons from the corona generates the reflection component
and the iron line.  However, recent observations and theoretical work
have shown that this geometry is not appropriate for explaining the
hard state spectra of Cyg X-1.  For example, the covering factor
inferred by Gierli\'nski et al. (1997a) is too small to be consistent
with the slab corona geometry.  Similarly, the observed Fe K$\alpha$
line is too narrow to allow the disk to extend close to the black hole
(Ebisawa et al. 1996).  More detailed theoretical modeling of the
standard disk-corona configuration show that models in which the disk
extends all the way to the last stable orbit do not allow a high
enough electron temperature in the corona to reproduce the observed
hard power-law slope and high energy cut-off (Dove et al. 1997a, 1997b;
Poutanen, Krolik \& Ryde 1997).  The logical conclusion is that the
thin disk is truncated at a radius equal to at least a few tens of
$R_{schw}$, and that the inner region of the accretion flow is filled
with hot gas.

Qualitatively, this conclusion is in perfect agreement with the
prediction of the EMN model for the low spectral state of accreting
black hole systems (see \S\ref{transition}).  Figure 3 shows that
there is also a reasonable quantitative agreement between the model
and the hard state spectra of Cyg X-1.  The data shown on the figure
correspond to nearly simultaneous Ginga and OSSE observations made on July 6
1991 (Gierli\'nski et al. 1997a).  For the spectral fitting we used a
grid of ADAF model spectra, characterized by three parameters: $\rtr$,
the HI column density along the line of sight to Cyg X-1, $N_{\rm H}$
(since the data does not extend below 3 keV, we did not obtain any
meaningful constraint on $N_{\rm H}$), and the overall normalization.
In each model we fixed $m=9$, $i=40^{\circ}$, $\beta=0.5$,
$\alpha=0.3$ and set the mass accretion rate to its critical value.

The best fitting model spectrum corresponds to $\rtr \sim 100$ and is
shown in Figure 3 as a solid line. Clearly, this model does a
reasonable job of reproducing both the spectral slope and the shape of
the exponential cut-off. The value of the transition radius derived
from this fit is consistent with the observed narrow width of the iron
fluorescence line (e.g.  Ebisawa et al. 1996).  To illustrate how
sensitive the spectrum is to the value of $\rtr$, we show the spectrum
corresponding to $\rtr =30$, keeping the other parameters the same
(dashed line).  This curve produces a slightly worse fit in the high
energy end of the spectrum, but the difference between the two model
spectra above 2 keV is no greater than $10\%$, comparable to the
uncertainty introduced by the approximations made in our current
calculations.  For example, using the relativistic global flow
solutions in the Kerr geometry (e.g.  Gammie \& Popham 1997) instead
of those based on Newtonian physics, will produce higher electron
temperatures in the ADAF (as demonstrated by Narayan et al. 1997d),
and therefore may well change the shape of the high energy cut-off.
We would therefore say that Cyg X-1 in the low state must have a
transition radius $\rtr \gsim 20$.

The exact value of the black hole mass does not affect the overall
shape of the X-ray spectrum above $\sim 2 $ keV, but the inclination
of the system does have a significant effect both on the shape and
normalization of the spectrum (e.g. see EMN).  Since the density
profile in an ADAF is slightly flattened (Narayan \& Yi 1995a), for
lower values of $i$ the observer sees deeper into the flow, i.e.  the
observed photons are produced in a hotter medium.  Moreover, because
of lower optical depth near the pole, the scattered photons escape
preferentially in that direction.  As a result, systems observed at
lower inclination are significantly brighter, and have higher
exponential cut-off energies.  We find that our fit above 100 keV
becomes significantly worse for $i \gsim 40^{\circ}$, which forces us
to adopt $m\gsim 9$, to be consistent with the other known binary
parameters.  However, with $m=9$ and $i = 40^{\circ}$ the model
spectrum shown on Figure 3 is $\sim 3$ times more luminous than the
data.  This normalization error is most likely the result of the
simplified way we have modeled the flattening of the accretion flow
(see \S\ref{dynamics} and Narayan et al. 1997a, EMN).  Because we have
truncated our model near the poles, the effect of changing the
inclination in the vicinity of $40^{\circ}$ is very dramatic, but
perhaps not very realistic.  In view of this limitation in the model,
as well as possible uncertainty in the distance to Cyg X-1, we feel
that the error in the normalization does not represent a serious
problem.

Our simplified treatment of the ADAF density profile is also one of
the causes for the slight discrepancy between our model spectra and
the data in the region around $10$ keV.  Figure 3 shows that the iron
K$\alpha$ line equivalent width and the depth of the corresponding Fe
edge predicted by the model fall short of what is observed, implying
that the covering factor of the cold disk is too small.  By allowing
the photons scattered in the ADAF to escape preferentially in the
direction away from the thin disk, the truncation of the hot flow near
the poles is at least partially responsible for this discrepancy.
Another effect we have not included is partial ionization of the gas
in the thin disk.  Since highly ionized iron ions have significantly
larger fluorescent yields than neutral atoms, taking into account the
ionization state of the thin disk may significantly increase the
predicted strength of the Fe line (e.g. $\dot{\rm Z}$ycki \& Czerny
1994).

\subsection{Hard--Soft and Soft--Hard Spectral Transitions}
\label{cygx1_transition}

Occasionally, Cyg X-1 undergoes a transition from the hard to the soft
spectral state, and the spectrum switches to one which is dominated by
an ultrasoft component with a temperature $\sim 0.3-0.4$ keV (Cui et
al. 1997a; Dotani et al. 1997). The spectrum above 10 keV becomes much
softer than in the hard state, with a variable photon index, $\Gamma
\sim 1.9-2.5$ (Gierli\'nski et al. 1997b; Cui et al. 1997a; Dotani et
al. 1997).

Since Cyg X-1 spends only a small fraction of time in the soft state,
this state is relatively less well studied compared to the hard state.
Recent observations (e.g. Zhang et al. 1997a; Cui et al. 1997a; Dotani
et al. 1997) during the state transition which occurred in the summer
of 1996 resulted in several important discoveries.  By combining
ASM/RXTE and BATSE lightcurves, Zhang et al. (1997a) demonstrated that
the total X-ray luminosity of Cyg X-1 remained practically constant
during the transition.  On the other hand, a comparison of the
characteristic temperature and luminosity of the ultrasoft blackbody
component before and during the transition showed that the apparent
inner radius of the thin disk was at least $\sim 3$ times larger in
the low state that in the soft state.  The authors concluded that the
change in the spectrum reflects a change in the relative importance of
the energy release in the hot and cold regions of the flow, without a
significant variation in the mass accretion rate.  (Note that in a
later paper, Zhang, Cui \& Chen (1997b) proposed that the change in
the inner radius of the disk may result from a temporary change
between a prograde and retrograde disk around a rotating black hole.)

The temporal properties of Cyg X-1 during the 1996 spectral state
transition (Cui et al. 1997c) also showed strong evolution.  Cui et
al. discovered that the observed time lag between the soft and hard
energy bands is roughly $\sim 10$ times smaller in the soft state than
in the hard state.  If this time lag is due to Compton scattering of
soft disk photons in the hot ``corona'', its magnitude must be
strongly correlated with the optical depth of the hot region.  The
authors conclude that their result requires the presence of a larger
hot ``corona'' in the hard state than in the soft state.  Cui et
al. (1997c) also argue that the observed change in the shape of the
power density spectrum strongly supports this interpretation.

The results outlined above fit in very well with the predictions of
the EMN model for the low--high state transition.  In this scenario,
the softening of the spectrum is due to a decrease in the transition
radius $\rtr$ between the ADAF zone and the thin disk, and therefore,
a decrease in the size and optical depth of the hot ADAF.  This is
precisely what is seen in Cyg X-1.  Moreover, the presence of the
intermediate state, during which $\rtr$ changed from its large low
state to its small high state value, was clearly inferred both from
the spectral and temporal evolution of the system (Cui et al. 1997a,
1997c; Belloni et al. 1996).

In Figure 4 we show combined PCA/HEXTE spectra taken on 1996 May 22,
roughly 10 days after the onset of the hard--soft state transition
(Cui et al. 1997a), together with our intermediate model spectrum
modified to take into account interstellar absorption.  As in fitting
the low state spectrum above, we used a grid of models constructed by
varying $\rtr$, while keeping $m=9$ and $i=40^{\circ}$.  The best fit
to the data was found with $N_H = 2.7 \times 10^{22}\ {\rm cm^{-2}}$
and $\rtr = 3.5$.  The overall agreement between the model spectrum
and the data is fairly good.  Our calculation reproduces both the
characteristic temperature of the soft disk emission and the shape of
the power-law component.  The only obvious discrepancy occurs in the
vicinity of Fe K$\alpha$ line.  As in the case of the low state, the
predicted line strength is too small, and it is clear that more
detailed modeling of both the ADAF and the thin disk is necessary in
order to accurately reproduce the observed line emission.  Though the
fitted value of $\rtr$ is very close to the marginal stable orbit,
the model requires that Cyg X-1 was still in the intermediate
state, consistent with the conclusions of Cui et al. (1997a, N.B. the
authors use the term ``settling period'' instead of intermediate
state) and Belloni et al. (1996).

Did Cyg X-1 reach the ``true'' high state during its 1996 state
transition?  Cui et al. (1997a, 1997c) argued that it did, based on
the fact that the shape of the power density spectra observed between
June and August of 1996 are very similar to those seen in Nova Muscae
in its high state (Miyamoto et al. 1993).  In the context of the EMN
model, this implies that during the low--high state transition $\rtr$
decreased all the way from its value in the low state ($\sim 100$, see
\S\ref{cygx1_low}) to the marginally stable orbit in the high state,
when the ADAF was confined to the corona above the disk.

To illustrate that this scenario captures the essential spectral
features of the Cyg X-1 state transition, we use the combined
simultaneous ASM/RXTE (1.3-12 keV) and BATSE/CGRO (20-600~keV) spectra
of Cyg X-1 taken between April and October 1996.  Each spectrum is
averaged over a single CGRO spacecraft pointing period (1-3 weeks),
except for a few very short pointings, for which the data from two
nearby periods are combined.  The ASM data points are converted to
fluxes by comparing the count rates with the average ASM Crab Nebula
count rates in each channel. The BATSE spectra are deconvolved by
assuming the simplest spectral model, namely a power-law with an
exponential cutoff.  In Figure 5(a) we plot the spectra taken during
the first half of the observation period, i.e. they should represent
the transition from the low to the high state.  The spectra taken
during the subsequent high--low transition are shown in Figure 5(b).
For comparison, our intermediate state spectra, computed by setting
$m=9$, $\mdot = \mdotcr \sim 0.11-0.12$ and varying $\rtr$ from $100$
to $3.2$, and a high state spectrum with $\rtr =3$, are plotted in
Figure 5(c).

By comparing the three panels in Figure 5, one can see the remarkable
similarity between the data and the EMN model. Our sequence of
intermediate and high state spectra clearly shows the three main
characteristics of the observed spectral evolution of Cyg X-1.  

(1) The softening of the spectral slope above 10 keV is very well
reproduced by the model, which gives $\Gamma \sim 1.5$ for the hardest
spectrum and $\Gamma \sim 2.3$ for the softest.  

(2) The model spectra clearly show the anti-correlation between the
fluxes in the soft and hard X-ray bands.  Moreover, the relative
normalization of the ultrasoft blackbody and the power-law component
is very close to what is observed.  

(3) According to the model, the bolometric luminosity changes by only
$\sim 50\%$ during the entire sequence.  This is in excellent
agreement with the observations (Zhang et al. 1997a).

(4) Finally, we clearly reproduce the characteristic spectral pivoting
around $\sim 10$ keV.

In the EMN model, the low--high state transition is triggered by a
slight increase of the mass accretion rate in the system.  For
transient BHXBs, the change in $\mdot$ is most likely due to a thermal
instability in the outer thin disk (e.g. Hameury et al. 1997).  What
causes the increase in the mass accretion rate in Cyg X-1?  Disk
instability is again a possible solution.  However, if $\mdot \sim
\mdotcr$ in the low state (as we have assumed in our discussion
above), we need very small changes in $\mdot$ to cause a significant
change in $\rtr$.  For instance the difference between the values of
$\mdot$ in the hardest and softest states in Figure 5c is only $\sim
15\%$.  Since in Cyg X-1, the mass transfer is powered by a wind from
the supergiant companion star (Gies \& Bolton 1986b), fluctuations in
$\mdot$ on the level of a few tens of percent are not improbable.

\section{GRO J0422+32} 
\label{J0422}
The transient source GRO J0422+32 was discovered in outburst by
BATSE/CGRO on August 4 1992 and reached its maximum luminosity in the
40-150 keV range by August 8 (Paciesas et al. 1992).  Its fast rise
was followed by an exponential decay with a characteristic time scale
$\sim 40$ days, a secondary maximum $\sim 140$ days after the onset of the
initial outburst (e.g. Harmon et al. 1992, 1994), and a steepening of
the decay after about 230 days.  The overall light curve resembled
closely those of typical transient sources like Nova Muscae, GS
2000-25 and A0620-00 (Tanaka \& Shibazaki 1996 and references therein).

GRO J0422+32 was observed at various times during the first few months
of the outburst by ASCA (Tanaka 1993), ROSAT (Pietsch et al. 1993),
Mir/TTM (Sunyaev et al. 1993), and OSSE (Grove et al. 1997).  Its
X-ray spectrum in all cases was hard and reasonably well-fitted by an
exponentially truncated power-law, similar to Cyg X-1 in its low
state.  An interesting characteristic of this transient is a complete
absence of an ultrasoft component during the outburst, similar to
another X-ray Nova, GS 2023-338 (Sunyaev et al. 1993).  The peak
luminosity in 2-300 keV is estimated to be $\sim 5\times 10^{37}\ {\rm
erg\,s^{-1}}$ (Sunyaev et al. 1993 scaled for $d=2.6$ kpc, see
\S\ref{J0422_param}), which corresponds to $\sim 0.05 L_{\rm Edd}$ for
an accreting object of mass $m=9$ (see \S\ref{J0422_param}). 

\subsection{Binary Parameters}
\label{J0422_param}
An optical counterpart of GRO J0422+32 with peak magnitude $V\sim
13.2$ was identified by Castro-Tirado et al. (1993).  Follow-up
observations showed that in quiescence the source dropped down to
$V=22.4$ (Zhao et al. 1994), representing a change of over 9 magnitudes.
Further study revealed that the system is a low-mass binary with an
orbital period $P=5.0$ hours and a mass function $f(M) = 1.21 \msun$
(Filippenko, Matheson \& Ho 1995; Casares et al. 1995a; Chevalier \&
Ilovaisky 1996).  

The optical spectrum during quiescence shows the secondary to be a
normal M2 dwarf with a mass $\sim 0.39 \msun$ (Filippenko et al. 1995;
Casares et al. 1995b).  The inclination of the system is highly
uncertain.  Filippenko et al. (1995) derived $i = 48^{\circ}$ from a
study of the H$\alpha$ emission line, which gives a black hole mass $M
= 3.6 \msun$.  Kato, Mineshige \& Hirata (1995) claimed to have
detected an eclipse feature in the orbital lightcurve, implying
that the binary is close to being edge-on.  Recently, however, Beekman
et al. (1997) studied the flux variation due to the distortion of the
secondary and concluded that $13^{\circ} \leq i \leq 31^{\circ}$,
assuming that $30\%$ of the light in the $I$-band comes from the
accretion disk (Filippenko et al. 1995).  This result implies that
the black hole mass has a lower limit of $\sim 9 \msun$, which is much
higher than previously thought.  In what follows we will adopt $m =
9$.

The distance to the binary can be calculated using the method
described by Barret, McClintock \& Grindlay (1996, the authors
identify it as ``method I''), in which they derive the absolute
magnitude of the secondary and compare it to the apparent de-reddened
magnitude in quiescence.  The luminosity of the secondary is estimated
using its known spectral type, which can be converted into a visual
flux (Popper 1980), assuming that the radius of the star is equal to
the average Roche lobe radius (Paczy\'nski 1971). We obtain $d \simeq
2.6$ kpc.  (This number is different from that quoted by Barret et al.
[1996] since they used a different visual extinction value; here we
adopt $A_V = 1.2$ [Filippenko et al. 1995]).  Note that this result
does not depend on the mass of the black hole or the inclination of
the system; the only relevant parameters are mass and spectral type of
the secondary, and its apparent visual magnitude.  Because of this, we
feel that our distance estimate is quite robust.

\subsection{Outburst Spectrum} 
\label{J0422_low}
The relatively low peak luminosity, the absence of a soft component,
and the general similarity of the spectrum of GRO J0422+32 to the low
state of Cyg X-1 (Sunyaev et al. 1993), lead us to propose that at the
peak of its outburst this system was in the low spectral state, as
described in \S\ref{states}.  In our scenario, the rapid rise and
decline of the X-ray luminosity occurred through a dramatic change in
the mass accretion rate, caused perhaps by an instability in the outer
thin disk (e.g. Cannizzo 1993; Hameury et al. 1997).  However, since
the spectrum of the system remained hard, and none of the published
spectra show any indication of the Compton reflection component or
iron fluorescence (e.g. see Figure 6), we conclude that the transition
radius between the hot and cold emission regions was at all times
greater than $\sim 100 R_{\rm Schw}$, and that the mass accretion rate
remained at or below its critical value, $\mdotcr \sim 0.1$.  Thus,
this outburst of GRO J0422+32 is significantly different from that of
Nova Muscae 1991, in which the accretion rate apparently reached the
Eddington limit and the system went from the quiescent to the very
high spectral state (EMN).  In GRO J0422+32, the transition occurred
between the quiescent and low spectral states.

In Figure 6 we plot a combined TTM (2-20 keV), HEXE (20-200 keV, from
Sunyaev et al. 1993 and Maisack et al. 1993) and OSSE (50-600 keV,
from Grove et al. 1997) spectrum of GRO J0422+32 taken three weeks
after the X-ray peak of the outburst (1992 August 29 -- September 2).
This period corresponds to approximately half maximum intensity at 100
keV.  Grove et al. fitted this data using an exponentially truncated
power-law model, and obtained a best fit photon index of 1.5 and
e-folding energy of $\sim 130$ keV.  Superimposed on the data are our
model spectra calculated for $m=9$, $\alpha=0.3$, $\beta = 0.5$,
$\mdot = 0.1$ and values of $i$ ranging from $15^{\circ}$ to
$75^{\circ}$, as shown in the figure.  We adopted $\rtr = 10^4$, but
the results are the same for other values of $\rtr$ down to $10^2$.

All three model spectra shown in Figure 6 reproduce the observed
spectral slope very well, though to fit the shape of the exponential
cut-off we need $i\gsim 40^{\circ}$.  Moreover, the spectrum computed with
$i=45^{\circ}$ gives the correct normalization (assuming d=2.6 kpc) as
well as spectral shape.  The binary parameters discussed in
\S\ref{J0422_param} require that, for a $9\msun$ black hole, the system
inclination must be around $31^{\circ}$.  The discrepancy between this
value and our estimate of $45^{\circ}$ highlights some of the
limitations of the model (see the discussion in \S\ref{cygx1_low}).

\subsection{E-folding Energy vs. Luminosity}
\label{J0422_temp}

OSSE observed GRO J0422+32 for 33 days from the peak of the outburst.
Throughout this period the spectrum was well described by an
exponentially truncated power-law.  However, Grove et al. (1997) found
that the best fit model parameters underwent a substantial evolution.
In particular, as the hard X-ray luminosity declined with time, the
e-folding energy increased nearly monotonically, i.e. the two
parameters were anti-correlated.  The nearly linear relation between
these two parameters is shown in Figure 7.  This result is in good
agreement with what is expected in the low spectral state (see Figure
2 and the discussion in \S\ref{transition}).  If, as we have assumed,
the mass accretion rate in GRO J0422+32 at the peak of the outburst is
near its critical value, and the decline of the hard X-ray flux is a
result of a decrease in $\mdot$, the temperature of the hot flow must
be anti-correlated with luminosity (the segment between low state and
quiescent state in Figure 2).

One must keep in mind, however, that a detailed comparison between
Figures 2 and 7 is difficult.  The e-folding energies plotted in
Figure 7 were obtained by fixing the photon index of the spectra to
$\Gamma = 1.5$, which is the best fit value for the combined spectrum
shown in Figure 6.  This was done because the OSSE data do not extend
to low enough energies to allow a meaningful fit of both $E_f$ and
$\Gamma$.  It is not clear from the data whether the photon index
changed during the period when the system was observed.  On the other
hand, our model predicts a $15\%$ change in the spectral slope during
this period, with the hardest spectra near the outburst maximum.
Though this discrepancy would not affect the overall qualitative
nature of our result, it would undoubtedly change the exact values of
$E_f$ for a given luminosity.  Moreover, the numerical values for the
gas temperature plotted in Figure 2 do not exactly correspond to the
e-folding energy of the observed spectrum (though we expect the two to
be proportional to each other), since $E_f$ is in a sense a measure of
the {\em average} electron temperature in the ADAF.

\section{GRO J1716$-$24}
\label{1716}

The transient system GRO J1719$-$24 was discovered in outburst on 1993
September 25 by BATSE and SIGMA (Harmon et al. 1993; Ballet et al.
1993).  The hard X-ray flux (20-100 keV) reached its maximum of $\sim
1.4$ Crab (e.g. van der Hooft 1996) five days after the original
detection, and for the next $70$ days the flux level remained high,
declining by no more than $\sim 20\%$.  This was followed by an abrupt
drop, during which the source flux dipped below the BATSE 1 day
detection limit of 0.1 Crab.  For the next $\sim 200$ days the flux
level remained undetectable by BATSE.  However, between September 1994
and September 1995, the light curve of the source exhibited five
separate flare events, with maximum amplitude roughly $1/3$ of the
original outburst (e.g.  Hjellming et al. 1996).  These secondary flares
were characterized by slow, almost linear rises and abrupt decays,
opposite to what is typically observed in X-ray novae (e.g. see a review
by Tanaka \& Shibazaki 1996). 

The optical counterpart for GRO J1719$-$24 was found by Della Valle,
Mirabele \& Cordier (1993).  From the interstellar absorption, Della
Valle, Mirabele \& Rodriuez (1994) estimated the distance to be $\sim
2.4$ kpc, which together with the upper limit on the brightness of the
system in quiescence, allowed them to conclude that the companion is a
main-sequence star of spectral type later than K.  Interpreting the
prominent optical period of $14^{\rm h}.7$ as a superhump phenomenon,
Masetti et al. (1996) estimated that the mass of the compact object
must be $M > 4.6 \msun$.  Thus, GRO J1719$-$24 appears to be a low-mass
black hole system.

On September 25-26 1993 the source spectrum was hard, similar to the
low state of Cyg X-1, with a power-law photon index $\Gamma = 2.0$ in
the hard X-ray (20-100 keV) band, and $\Gamma = 1.6$ in the softer
(2-27 keV) band (Harmon et al. 1993; Kaniovsky, Borozdin, \& Sunyaev
1993).  Van der Hooft et al. (1996) report that the 20-100 keV
spectrum softened to $\Gamma = 2.3$ during the rise to the first
maximum and continued to soften gradually thereafter. 

The spectral behavior of GRO J1719$-$24 during the secondary flares (as
we will denote the five smaller events to distinguish them from the
original outburst) in 1994-95 is rather uncertain.  OSSE observations
at the trailing edge of the first flare (1994 November $9-14$) showed a
power-law spectrum with no high energy cut-off and photon index
$\Gamma \sim 2.4$ (Grove et al. 1997).  The same team observed a hard
exponentially truncated power-law spectrum with $\Gamma \sim 1.5$ near
the peak of the second flare (1995 February 1-14). Just two days
later, the source was detected by the MIR/Kvant team, who reported a
power-law spectrum (in the energy band 2-27 keV) with a photon index
$2.1 \pm 0.3$.

To summarize, it appears that GRO J1719$-$24 was observed in three
distinct spectral states, two of which are similar to the soft and
hard states of Cyg X-1.  When the source was bright in the hard X-ray
band during the flares, its spectrum was best characterized as a hard
power-law with an exponential cut-off at $\sim 100-130$ keV (low
state), while the periods of lower hard X-ray flux during the flares
seemed to be associated with the appearance of a steeper power-law
component with $\Gamma > 2.0$ and no cut-off below $\sim 500$ keV
(high state).  The third spectral state appeared during the first and
largest outburst, and is characterized by high hard X-ray flux and a rather
soft spectrum with $\Gamma \lsim 2.3$, reminiscent of the very high
state seen in Nova Muscae and GX 339$-$4 (e.g. Ebisawa et al.  1994;
Miyamoto et al. 1991).

How can this behavior be interpreted in terms of the EMN spectral
state model?  Since the rise to peak intensity during the first
outburst was very similar to the outbursts exhibited by other black
hole transients (e.g.  Nova Muscae, A0620-00), we speculate that
during this time the system made a rapid transition from the quiescent
to the very high spectral state, following a dramatic increase in the
mass accretion rate. The 20--100 keV flux from GRO J1719$-$24 peaked
at roughly $\sim 0.4\,{\rm phot\,cm^{-2}\,s^{-1}}$ (van der Hooft et
al. 1996), which corresponds to $\sim 1.5\times 10^{37}\,{\rm
erg\,s^{-1}}$ (for a source with a photon index $\Gamma = 2.3$ at a
distance of 2.4 kpc), comparable to the luminosity of Nova Muscae in
the very high state (Ebisawa et al. 1994; EMN).  The rapid drop in the
hard X-ray flux $\sim 70$ days after the peak luminosity, is also very
similar to that seen in the lightcurve of Nova Muscae (EMN), and we
interpret it as the transition from the very high to the high state.
After that, the hard X-ray flux remained below the BATSE threshold
until the accretion rate dropped low enough for an ADAF to form,
i.e. the system entered an intermediate state and moved into the low
state.

The most straight-forward interpretation of the data during the
secondary flares directly follows from Grove et al. (1997).  At the
time of flare maxima, the system was in the low (or perhaps
intermediate) spectral state, characterized by a hard power-law
spectrum with a thermal cut-off.  Unfortunately, the data are not good
enough to place constraints on the presence of the Compton reflection
component, so that the position of the transition radius is uncertain,
though it is likely that $\rtr \gsim 30$.  We propose that the rapid
decay which accompanied each secondary flare was caused by an increase
in $\mdot$ which precipitated a corresponding decrease in $\rtr$, or a
disk instability of some sort which caused $\rtr$ to decrease
suddenly.  Thus, during the times when the flux in the BATSE band was
low, we propose that the system was in the high spectral state, with
$\rtr$ close to the last stable orbit.

In Figure 8 we plot the e-folding energy obtained from the OSSE data
taken on 1995 February 1-14 near the peak of the second flare, against
the integrated OSSE flux.  During this time, the X-ray luminosity was
decreasing nearly monotonically with time.  As in Figure~6, the values
of $E_f$ were derived using an exponentially truncated power-law model
with a fixed photon index, $\Gamma=1.5$.  We see that the hard X-ray
luminosity and the e-folding energy are correlated.  A comparison with
Figure 2 shows that this behavior is expected if the system is in the
intermediate state and is switching from the low to the high state.
Thus, the data in Figure 8 confirms the proposal we have made for the
secondary flares.  

A detailed comparison of Figures 7 and 8 shows that the values of
$E_f$ in GRO J1719$-$24 are somewhat higher compared to GRO J0422+32.
According to the model, the minimum $E_f$ in Figure 7 must be roughly
equal to the maximum $E_f$ in Figure 8 (see Figure 2).  Instead the
latter appears to exceed the former by $\sim 30$ keV, though the
errorbars are large.  If this result is real it might imply that GRO
J1719$-$24 has smaller inclination, or perhaps larger black hole spin.

Since during the high spectral state most of the system luminosity
comes out as ultrasoft blackbody disk emission, observations in the
energy range below 10 keV would provide the most direct test of our
interpretation of the data on GRO J1719$-$24.  Unfortunately, we are not
aware of any such observations made during the times when the source
was ``off'' in the BATSE energy band.  According to our
interpretation, the system should have been in the high state and,
therefore, very bright around 1 keV.

In conclusion, we emphasize that the behavior of GRO J1719$-$24 is
qualitatively quite similar to that of Nova Muscae 1991.  Both sources
showed a rapid increase in the 20-100 keV flux, followed by an abrupt
drop, and followed later by a secondary maximum.  However, there are
two important differences between the lightcurves of the two systems.
Firstly, the complete outburst in GRO J1719$-$24 lasted at least twice
as long as that in Nova Muscae, indicating that the time scale for the
decrease of the mass accretion rate was different in the two systems.
Secondly, GRO J1719$-$24 showed five secondary flares, instead of just
one observed in Nova Muscae.  According to our interpretation, these
flares were caused by oscillations of the transition radius, either as
a result of fluctuations in the mass accretion rate or because of a
disk instability.  Nova Muscae, perhaps because of the mare rapid
decline of $\mdot$, did not undergo such oscillations.

\section{Conclusions}
\label{discuss}

The ADAF solution (Ichimaru 1977; Rees et al. 1982; Narayan \& Yi
1994, 1995b; Abramowicz et al. 1995; Chen et al. 1995) is currently
the only known stable accretion model that explains the observed X-ray
spectra of BHXBs with a self-consistent treatment of gas dynamics and
radiative transfer.  The model satisfies mass and momentum
conservation at each radius, solves for the electron and ion
temperatures through the use of appropriate energy equations, and
includes a fairly sophisticated treatment of radiative processes.
Narayan (1996) and EMN used the ADAF solution to develop a unified
picture of the quiescent, low, intermediate and high spectral states
observed in BHXBs.  In their model, the accretion flow consists of two
distinct parts, a hot ADAF inside some transition radius $\rtr$ (in
Schwarzschild units), and a standard thin disk with a hot corona 
outside $\rtr$.  The different spectral states then correspond to
different values of the two main parameters, $\mdot$ and $\rtr$.

The basic scenario that emerged from the work of Narayan (1996) and
EMN is summarized in \S2.  In the quiescent and low spectral states
the transition radius is relatively large, $\rtr \gsim 100$, and
variations in the observed spectrum result solely from changes in the
mass accretion rate, $\mdot$. In the quiescent state, we have $\mdot <
10^{-3}$, while in the low state $\mdot$ can be as high as about 0.1
(in Eddington units).  The low state survives only until $\mdot$
reaches a critical value $\mdotcr \sim 0.1$.  When $\mdot > \mdotcr$,
the outer parts of the ADAF can no longer be in thermal equilibrium
and the cool disk begins to encroach into the ADAF.  When the
transition radius lies between $\rtr \sim 100$ and 3 (with $\mdot \sim
\mdotcr$) we have the so-called intermediate state.  Finally, above a
certain $\mdot$, $\rtr$ comes down to the marginally stable orbit,
$\rtr = 3$, and the ADAF is restricted to a corona above the
disk. This is the high state.

Some of the basic features of the EMN model are similar to other
models that have appeared previously in the literature.  Shapiro et
al. (1976) and Ichimaru (1977) suggested that the low state of Cyg X-1
correspond to an optically thin, hot accretion flow, rather than the
standard thin disk.  Ichimaru further speculated that the transition
between the hard and soft states in Cyg X-1 might be caused by a
change in the outer boundary conditions of the accretion flow, which
determines whether the accreting gas assumes the standard thin disk
configuration or forms a hot optically thin flow.  However, the model
proposed by EMN is the first successful attempt to unify the
quiescent, low, intermediate and high spectral states within the
framework of a single self-consistent calculation (see Figure 1).

Application of the model outlined above to the outburst of Nova Muscae
1991 (EMN) demonstrated that it reproduces well the individual low,
intermediate and high state spectra of this system.  Moreover, the
theoretical light curve based on a simple mapping between the mass
accretion rate, transition radius and time is in good agreement with
the observed X-ray and $\gamma$-ray light curves.  Further
confirmation of the main results of EMN came from more detailed
analysis of the spectra of Nova Muscae observed while the system was
in the intermediate state ($\dot{\rm Z}$ycki et al. 1997).  From
detailed fitting of the Fe K$\alpha$ emission, $\dot{\rm Z}$ycki et
al. concluded that the inner edge of the thin disk did increase during
this period, as predicted by EMN.  $\dot{\rm Z}$ycki et al. also
showed that there is a discrepancy in the exact value of $\rtr$ during
the intermediate state, but this is not a serious problem, as we have
argued in \S2.2.

The present paper describes further tests of the EMN model on other
black hole systems.  In \S3 we analyze data taken during the 1996
hard--soft state transition of Cyg X-1 and compare with the
predictions of the model.  The results are encouraging.  The hard and
soft spectra of Cyg X-1 are reasonably well fitted by low state and
intermediate state model spectra (Figures 3 and 4).  The best fit
value of the transition radius for the hard state is $\rtr \sim 100$,
consistent with the narrow width of the iron fluorescence line
observed by Ebisawa et al. (1996) in the low state.  The spectrum
corresponding to $\rtr=30$ is somewhat too soft to reproduce the data,
though we feel that the uncertainties in our model (e.g. use of
Newtonian dynamics and radiative transfer, as well as a simplistic
treatment of the flattening of the ADAF density profile) are
comparable to the difference between those two curves.  We would thus
say that the low state in Cyg X-1 is consistent with any $\rtr \gsim
20$.  The soft state spectrum is best described by a model with $\rtr
\sim 3.5$, i.e. close to the marginally stable orbit.

We have also compared a sequence of intermediate and high state
spectra computed for several values of $\rtr$ between 100 (its value
in the low state) and 3 (the marginally stable orbit for a
non-rotating black hole), to observations made by BATSE and RXTE/ASM
during the 1996 transition in Cyg X-1.  From Figure 5, it is clear
that in addition to explaining the broad qualitative features of the
transition, the model reproduces very well several details seen in the
observations.  The range of photon indices obtained with the model,
$\Gamma \sim 1.5-2.3$, is very similar to that seen in the data. The
model and the data both show anti-correlation between the soft and
hard X-ray flux.  The model also reproduces the pivoting around 10 keV
seen in the data as well as the nearly constant bolometric luminosity
throughout the transition.  The striking similarity between the model
and observations indicates that the general picture (Narayan 1996,
EMN) in which the low--high state transition arises from a change in
the transition radius (perhaps as a result of a minor fluctuation in
the mass accretion rate) must be close to the truth.

In \S2.3 we review the main spectral signatures of the state
transitions predicted by the EMN model, and discuss how they can be
used to deduce the physical changes in the accretion flow that
accompany such transitions.  We show that in both the low state and in
the intermediate state a decrease in hard X-ray luminosity must
invariably be accompanied by a softening of the overall spectrum.
Thus, from the point of view of an observer, if all one has is hard
X-ray data, it is difficult to distinguish between a decrease of
$\mdot$ in the low state and a decrease in $\rtr$ in the intermediate
state.  We show, however, that the position of the high energy cut-off
is a much better diagnostic.  In the low state, the e-folding energy
and the hard X-ray luminosity are anti-correlated, since at lower mass
accretion rates the ADAF is hotter.  The exact opposite happens in the
intermediate state, where a decrease in $\rtr$ causes both the hard
X-ray flux and the temperature of the hot flow to drop.  The two
effects are illustrated in Figure 2. To show how these properties of
the model could be exploited in interpreting observations of real
systems, in \S\S4 and 5 we discuss the transient systems GRO J0422+32
and GRO J1719$-$24.

We find that the observed spectral evolution of GRO J0422+32 after the
peak of its 1992 outburst must have corresponded to a transition
between the low and quiescent spectral states.  This interpretation
arises from the fact that the cut-off temperature was anti-correlated
with the X-ray luminosity during the first 30 days of the decline when
OSSE observed the source (Figure 7).  Moreover, we find that the
spectrum observed at roughly half maximum luminosity is very well
fitted by a low state model spectrum computed with $m=9$, and $\rtr =
10^4$.  Thus, GRO J0422+32 seems to have been very different from a
more ``typical'' transient like Nova Muscae or A0620$-$00.  In GRO
J0422+32 it would appear that the mass accretion rate even at the peak
of the outburst was at or below the critical rate $\mdotcr$,
i.e. $\Mdot \lsim 0.1 \Mdot_{\rm Edd}$.  Thus, GRO J0422+32 was
dominated by an ADAF all through its outburst. Interestingly, the
accretion rate in this system in quiescence is estimated at $\mdot\sim
10^{-4}$ (Menou 1997, private communication), significantly lower than
quiescent values in other transient systems (Narayan et al. 1996,
1997a).

On the other hand, the outburst of GRO J1719$-$24 appears to have been
closer to that of Nova Muscae.  We make a tentative proposal that the
first and largest peak corresponds to a transition from the quiescent
to the very high state, characterized by high X-ray luminosity and a
power-law spectrum with photon index $\gsim 2.3$ (Hooft et al. 1996).
The sharp drop in the BATSE flux two month after the initial outburst
is very similar to that observed in Nova Muscae, which EMN identified
as the transition from the very high to the high state.  According to
our proposal, the secondary flares correspond to a series of
transitions between the high and intermediate spectral states.  Both
the shapes of the spectra and the observed correlation between the
e-folding energy and X-ray flux support this interpretation.  Of
course, the fact that GRO J1719$-$24 oscillated five times between the
high and intermediate state differentiates this source from Nova
Muscae and A0620$-$00 which had only one such transition.  Perhaps the
decline of $\mdot$ was not monotonic in GRO J1719$-$24, or more likely
the outer thin disk underwent some kind of a disk instability.

Although we have been quite successful in applying the EMN model to
Cyg X-1, GRO J0422+32 and GRO J1719$-$24, we would like to emphasize
that the main results presented in this paper are still somewhat
qualitative, and the uncertainties in the model are still too large to
draw meaningful quantitative conclusions.  The accretion flow
calculations employed here take into account most of the relevant
physical processes in a self-consistent way, and are superior to other
analyses in the literature.  Nevertheless, a number of improvements
need to be made before we can attempt detailed comparisons between our
models and the observational data.

The main avenue of future work lies in incorporating relativistic
effects into the calculations.  Relativistic global flow solutions in
Kerr geometry (Abramowicz et al. 1996; Peitz \& Appl 1997; Gammie \&
Popham 1997) and gravitational redshift have already been used in
modeling the spectrum of Sgr~A$^*$ (Narayan et al. 1997d).  In
calculating radiative transfer in an ADAF, we need to include the
effects of ray bending and Doppler boosts due to the motion of the gas
(Jaroszynski \& Kurpiewski 1997), and to modify our treatment of
Compton scattering to take into account the bulk radial motion of the
gas (Titarchuk et al. 1996, 1997).  Our treatment of the thin disk
spectrum must also be modified to include relativistic effects.  The
strength and shape of the iron fluorescence line and the corresponding
absorption edge have recently been used as a successful diagnostic to
probe the geometry and physical characteristics of the accretion flow
(e.g. $\dot{\rm Z}$ycki et al. 1997; Cui et al. 1997b).  However, in
order to use these features to constrain our model we must assume a
more realistic density profile of the ADAF in the vertical direction,
instead of truncating the flow near the poles.  We also need to
compute self-consistently the ionization state of the thin disk.

\acknowledgements

We thank M. Gierli\'nski for providing Cyg X-1 low state data, and J.
Nevelainen, T. Aldcroft and J. McClintock for their help with XSPEC.
This work was supported in part by NSF grant AST 9423209 and NASA
grant NAG 5-2837.  A.~A.~E. was supported by a National Science
Foundation Graduate Research Fellowship.

\vfill\eject
\references
\def\refpar{\hangindent=3em\hangafter=1}
\def\reference{\refpar\noindent}
\def\apj{ApJ}
\def\apjs{ApJS}
\def\mnras{MNRAS}
\def\aa{A\&A}
\def\aas{A\&A Suppl. Ser.}
\def\aj{AJ}
\def\araa{ARA\&A}
\def\nat{Nature}
\def\pasj{PASJ}

\reference Abramowicz, M. A., Chen, X., Grantham, M., Lasota, J.-P. 1996, 
\apj, 471, 762

\reference Abramowicz, M. A., Chen, X., Kato, S., Lasota, J. P., \& Regev, O.
1995, \apj, 438, L37

\reference Ballet, J., Denis, M., Gilfanov, M., \& Sunyaev, R. 1993, IAU Circ. 
5874

\reference Balucinska-Church, M., Belloni, T., Church, M. J., Hasinger, G.
1995, \aa, 302, L5

\reference Barr, P., White, N. E., Page, C. G. 1985, \mnras, 216, 65

\reference Barret, D, McClintock, J. E. \& Grindlay, J. E. 1996, \apj,
473, 963

\reference Beekman, G., Shahbaz, T., Naylor, T., Charles, P. A.,
Wagner, R. M., \& Martini, P. 1997, \mnras, 290, 303

\reference Belloni, T., Mendez, M., van der Klis, M., Hasinger, G., 
Lewin, W. H. G., van Paradijs, J. 1996, \apj, 472, L107 

\reference Bisnovatyi-Kogan, G. S. \& Lovelace, R. V. E. 1997, \apj,
486, 43

\reference Blackman 1997, submitted to Phys Rev Let  

\reference Bowyer, S., Byram, E. T., Chubb, T. A., Friedman, M. 1965, Science, 
147, 394

\reference Bregman, J., Butler, D., Kemper, E., Koski, A., Kraft R. P., 
Stone, R. P. S. 1973, \apj, 185, L117

\reference Cannizzo, J. K. 1993, in Accretion Disks in Compact Stellar 
Systems, ed. J. C. Wheeler, Singapore: World Scientific Publishing, 6

\reference Casares, H., Marsh, T. R., Charles, P. A., Martin, A. C., 
Martin, E. L., Harlaftis, E. T., Pavlenko, E. P., Wagner, R. M. 1995a, 
\mnras, 274, 565

\reference Casares, H., Martin, A. C., Charles, P. A., Martin, E. L., 
Rebolo, R., Harlaftis, E. T., Castro-Tirado, A. J. 1995b, \mnras, 276, L35

\reference Castro-Tirado, A. J., Pavlenko, E. P., Shlyapnikov, A. A., 
Brandt, S., Lund, N., Ortiz, J. L.  1993, \aa, 276, L37

\reference Chen, X. 1995, \mnras, 275, 641

\reference Chen, X., Abramowicz, M. A., Lasota, J. P., Narayan, R., Yi, I.
1995, \apj, 443, L61

\reference Chevalier, C. \& Ilovaisky, S. 1996, \aa, 312, 105

\reference Cui, W., Heindl, W. A., Rothschild, R. E., Zhang, S. N., 
Jahoda, K., Focke, W. 1997a, \apj, 474, L57

\reference Cui, W., Ebisawa, K., Dotani, T., \& Kubota, A. 1997b,
{\apj}L, in press

\reference Cui, W., Zhang, S. N., Focke, W., Swank, J. H. 1997c, \apj, 484, 383

\reference Della Valle, M., Mirabele, F., \& Cordier, B. 1993, IAU Circ. 5876

\reference Della Valle, M., Mirabele, I. F. \& Rodriuez, L. F. 1994, \aa, 290, 
803

\reference Dolan, J. F. 1992, \apj, 384, 249 

\reference Done, C., Mulchaey, J. S., Mushotzky, R. F., Arnaud, K. A. 1992, 
\apj, 395, 275

\reference Dotani, T. et al. 1997, \apj, 485, L87

\reference Dove, J. B., Wilms, J., Maisack, M., Begelman, M. C. 1997a,
\apj, 487, 759

\reference Dove, J. B., Wilms, J., Nowak, M. A., Vaughan, B. A., \&
Begelman, M. C. 1997b, submitted to \apj Letters

\reference Ebisawa, K. et al. 1994, \pasj, 46, 375

\reference Ebisawa, K., Titarchuk, L., \& Chakrabarti, S. K. 1996,
\pasj, 48, 59

\reference Ebisawa, K., Ueda, Y., Inoue, H., Tanaka, Y., White, N. E. 1996, 
\apj, 467, 419

\reference Esin, A. A., McClintock, J. E., \& Narayan, R. 1997, \apj, 489, 000

\reference Fabian, A. C., Rees, M. J., Stella, L., White, N. E. 1989, \mnras, 
238, 729

\reference Filippenko, A. V., Matheson, T., Ho, L. C. 1995, \apj, 455, 614

\reference Frank, J., King, A., \& Raine, D. 1992, Accretion Power in
Astrophysics (Cambridge, UK: Cambridge University press)

\reference Gammie, C. \& Popham, R. 1997, \apj, in press

\reference George, I. M. \& Fabian, A. C. 1991, \mnras, 249, 352

\reference Gierli\'nski, M., Zdziarski, A. A., Done, C., 
Johnson, W. N., Ebisawa, K., Ueda, Y., Haardt, F., Phlips, B. F. 1997a, 
\mnras, 288, 958

\reference Gierli\'nski, M., Zdziarski, A. A., Dotani, T., Ebisawa,
K., Jahoda, K., Johnson W. N. 1997b, in Proceedings of 4th Compton
Symposium, in press

\reference Gies, D. R. \& Bolton, C. T. 1986a, \apj, 304, 371

\reference Gies, D. R. \& Bolton, C. T. 1986b, \apj, 304, 389

\reference Gies, D. R. \& Bolton, C. T. 1982, \apj, 260, 240

\reference Gilfanov, M. et al. 1993, \aas, 97, 303

\reference Grove, J. E., Johnson, Kroeger, R. A., McNaron-Brown, K., 
Skibo, J. G., \& Phlips, B. F. 1997, submitted to \apj

\reference Gruzinov 1997, submitted to \apj

\reference Haardt, F., Done, C., Matt, G., Fabian, A. C. 1993, \apj, 411, L95

\reference Hameury, J.-M., Lasota, J.-P., McClintock, J. E., \& Narayan, R.
1997, 489, 234

\reference Harmon, B. A. et al. 1992, IAU Circ. 5685 

\reference Harmon, B. A., Zhang, S. N., Paciesas, W. S., \& Fishman, G. J. 
1993, IAU Circ. 5874

\reference Harmon, B. A. et al. 1994, in the Second Compton Symposium, eds. 
C. E. Fichtel, N. Gehrels, \& J. P. Norris (AIP, New York), p. 210

\reference Hawley, J. F. \& Balbus, S. A. 1996, in Physics of Accretion Disks,
eds. S. Kato et al., Gordon and Breach Sci. Publ., 273

\reference Herrero, A., Kudritzki, Gabler, R., Vilchez, J. M., Gabler, A. 
1995, \aa, 297, 556

\reference Hjellming, R. M., Rupen, M. P., Shrader, C. R., Campbell-Wilson, D., 
Hunstead, R. W., \& McKay, D. J. 1996, \apj, 470, L105

\reference Huang, M. \& Wheeler, J. C. 1989, \apj, 343, 229

\reference Ichimaru, S. 1977, \apj, 214, 840

\reference Jaroszy\'nski, M. \&  Kurpiewski, A. 1997, \aa, 326, 419 

\reference Kaniovsky, A., Borozdin, K., \& Sunyaev, R. 1993, IAU Circ. 5878

\reference Kato, T., Mineshige, S., \& Hirata, R. 1995, \pasj, 47, 31

\reference Lasota, J. P., Narayan, R., \& Yi, I. 1996b, \aa, 314, 813

\reference Liang, E. 1997, Phys. Rep., in press 

\reference Lightman, A. P. \& White, T. R. 1988, \apj, 335, 57

\reference Mahadevan, R., Narayan, R., Yi, I. 1996, \apj, 465, 327

\reference Maisack et al. 1993, in Second Compton Symposium, eds. 
C. E. Fichtel, N. Gehrels, \& J. P. Norris (AIP, New York), p. 304

\reference Margon, B. \& Bowyer, S. 1973, \apj, 185, L113

\reference Masetti, N., Bianchini, A., Bonibaker, J., Della Valle, M., \& 
Vio, R. 1996, \aa, 314, 123

\reference McClintock, J. E., Horne, K., \& Remillard, R. A. 1995, \apj, 
442, 358 

\reference Melia, F. \& Misra, R. 1993, \apj, 411, 797

\reference Mendez, M. \& van der Klis, M. 1997, \apj, 479, 926

\reference Meyer, F. \& Meyer-Hofmeister, E. 1994, \aa, 288, 175

\reference Mineshige, S. 1996, \pasj, 48, 93

\reference Mineshige, S. \& Wheeler, J. C. 1989, \apj, 343, 241 

\reference Miyamoto, S., Iga. S., Kitamoto, S., \& Kamado, Y. 1993,
\apj, 403, L39

\reference Miyamoto, S., Kimura, K., Kitamoto, S., Dotani, T., \& Ebisawa, K. 
1991, \apj, 383, 784

\reference Morrison, R. \& McCammon, D. 1983, \apj, 270, 119

\reference Narayan, R. 1997, in Proc. IAU Colloq. 163 on Accretion 
Phenomena \& Related Outflows, ASP Conf. Series, eds. D. T. Wickramasinghe
et al., 75

\reference Narayan, R. 1996, \apj, 462, 136

\reference Narayan, R., Barret, D., \& McClintock, J. E. 1997a, \apj, 482, 448

\reference Narayan, R., Garcia, M. R. \& McClintock, J. E. 1997b, \apj, 
478, 79

\reference Narayan, R., Kato, S. \& Honma, F. 1997c, \apj, 476, 49

\reference Narayan, R., Mahadevan, R., Grindlay, J. E., Popham, B.,
Gammie, C. 1997d, \apj, 492, in press

\reference Narayan, R., McClintock, J. E., \& Yi, I. 1996, \apj, 457, 821 

\reference Narayan, R. \& Yi, I. 1994, \apj, 428, L13

\reference Narayan, R. \& Yi, I. 1995a, \apj, 444, 231

\reference Narayan, R. \& Yi, I. 1995b, \apj, 452, 710

\reference Narayan, R., Yi, I., \& Mahadevan, R. 1995, \nat, 374, 623

\reference Ninkov, Z., Walker, G. A. H., Yang, S. 1987, \apj, 321, 425

\reference Novikov, I. D. \& Thorne, K. S. 1973, in Black Holes, ed. 
DeWitt, C. and B. (Gordon \& Breach, NY), 343

\reference Nowak, M. A. 1995, \pasp, 107, 1207

\reference Paciesas et al. 1992, IAU Circ. 5580

\reference Paczy\'nski, B. 1971, \araa, 9183 

\reference Peitz, J. \& Appl, S. 1997, \mnras, 286, 681

\reference Phlips, B. F. et al. 1996, \apj, 465, 907 

\reference Pietsch, N. et al. 1993, \aa, 273, L11

\reference Popper, D. M. 1980, \araa, 18, 115

\reference Poutanen, J., Krolik, J. H., \& Ryde, F. 1997, \mnras, in
press

\reference Quataert 1997, submitted to \apj

\reference Rees, M. J., Begelman, M. C., Blandford, R. D., \& Phinney,
E. S.  1982, \nat, 275, 17

\reference Shakura, N. I. \& Sunyaev, R. A. 1973, \aa, 24, 337

\reference Shapiro, S. L., Lightman, A. P., \& Eardley, D. M. 1976, \apj, 204,
187 (SLE)

\reference Shimura, T. \& Takahara, F. 1995, \apj, 445, 780

\reference Sunyaev, R. A. et al. 1993, \aa, 280, L1

\reference Tanaka, Y. 1993, IAU Circ. 5851 (J0422)

\reference Tanaka, Y. \& Lewin, W. H. G. 1995, X-ray Binaries, eds. W. H. G. 
Lewin et al., Cambridge Univ. Press., 126 

\reference Tanaka, Y. \& Shibazaki, N. 1996, \araa, 34, 607

\reference Titarchuk, L. G., Mastichiadis, A. \& Kylafis, N. D. 1996, \aas, 
120, C171

\reference Titarchuk, L. G., Mastichiadis, A. \& Kylafis, N. D. 1997,
\apj, 487, 834

\reference van der Hooft, F. et al. 1996, \apj, 458, L75
        
\reference van der Klis, M. 1994, \apjs, 92, 511

\reference Walborn, N. R. 1973, \apj, 179, L123

\reference White, T. R. Lightman, A. P. \& Zdziarski, A. A. 1988, \apj, 
v331, 939

\reference Zhang, S. N., Cui, W., Harmon, B. A., Paciesas, W. S., 
Remillard, R. E., van Paradijs, J. 1997a, \apj, 477, L95

\reference Zhang, S. N., Cui, W., Chen, W. 1997b, \apj, 482, 155

\reference Zhao, P. et al. 1994, IAU Circ. 5929 

\reference $\dot{\rm Z}$ycki, P. T., Done, C., \& Smith D. A. 1997, submitted 
to {\apj}L.

\reference $\dot{\rm Z}$ycki, P. T., Czerny, B. 1994, \mnras, 266, 653

\vfill\eject
\begin{figure}
\includegraphics{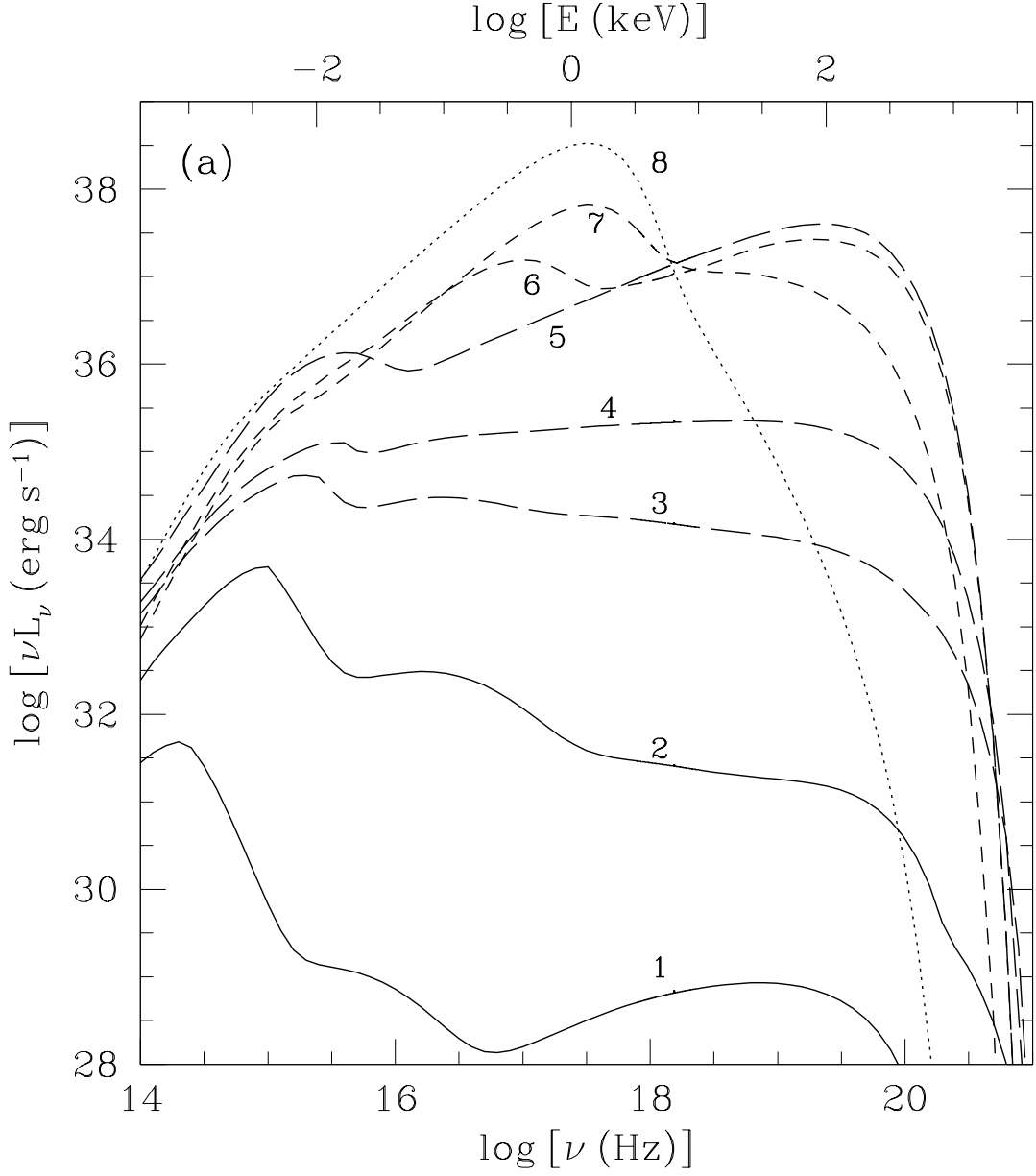} 
\vskip 6.5in 
\caption{(a) A sequence of spectra corresponding to the quiescent
state (solid line), low state (long-dashed line), intermediate state
(short-dashed line), and high state (dotted line), computed with
$m=9$, $i=40^{\circ}$, $\alpha=0.3$, $\beta=0.5$, and the following
values of $\log (\rtr)$ and $\mdot$: (a) 1--(3.9, $10^{-4}$); 2--(3.9,
$10^{-3}$); 3--(3.9, $0.01$); 4--(3.9, $0.03$); 5--(3.9, $0.11$);
6--(1.5, $0.11$); 7--(0.5, $0.12$); 8--(0.5, $0.4$).}
\end{figure}

\setcounter{figure}{0}

\begin{figure}
\includegraphics{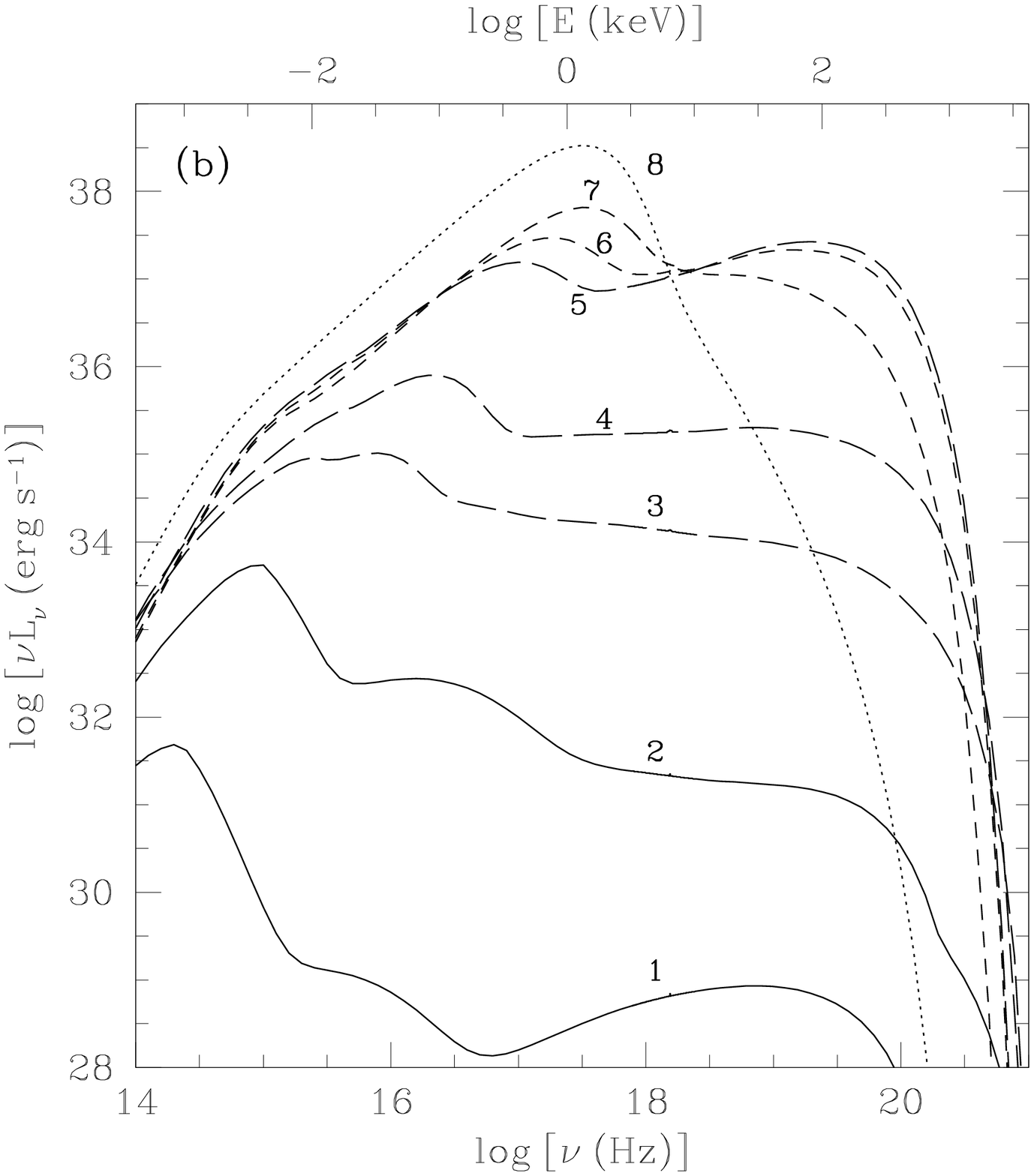} 
\vskip 6.5in 
\caption{(b) 1--(3.9, $10^{-4}$); 2--(3.2, $10^{-3}$); 3--(2.3,
$0.01$); 4--(1.9, $0.03$); 5--(1.5, $0.11$); 6--(1.0, $0.12$);
7--(0.5, $0.12$); 8--(0.5, $0.4$).}

\end{figure}

\begin{figure}
\includegraphics{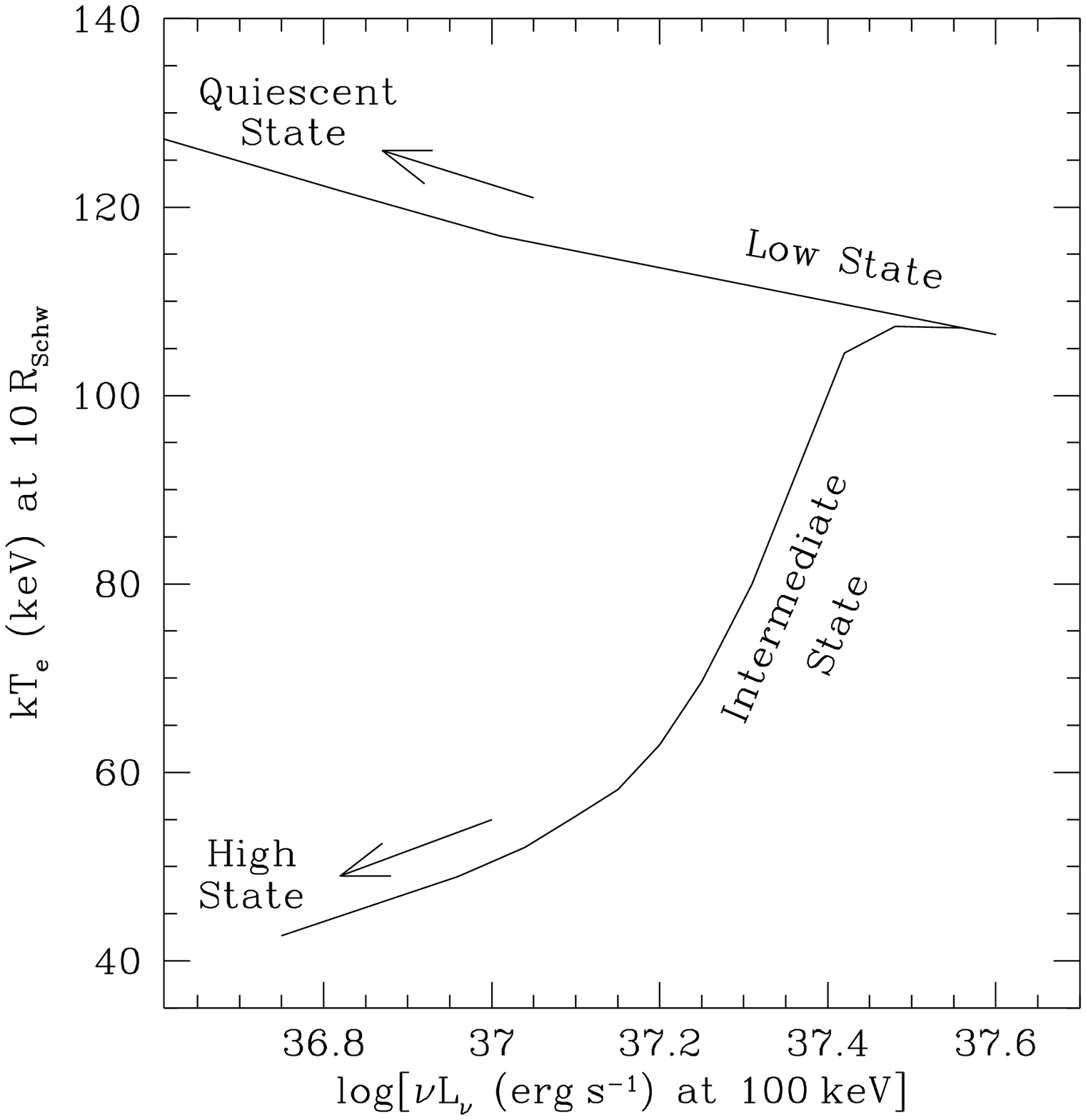} 
\vskip 6.5in 
\caption{The electron temperature in an ADAF at $10 R_{\rm Schw}$
plotted vs. the specific luminosity at 100 keV.  Note that the two
quantities are correlated in the intermediate and high spectral states
and anticorrelated in the quiescent and low states.}
\end{figure}

\begin{figure}
\includegraphics{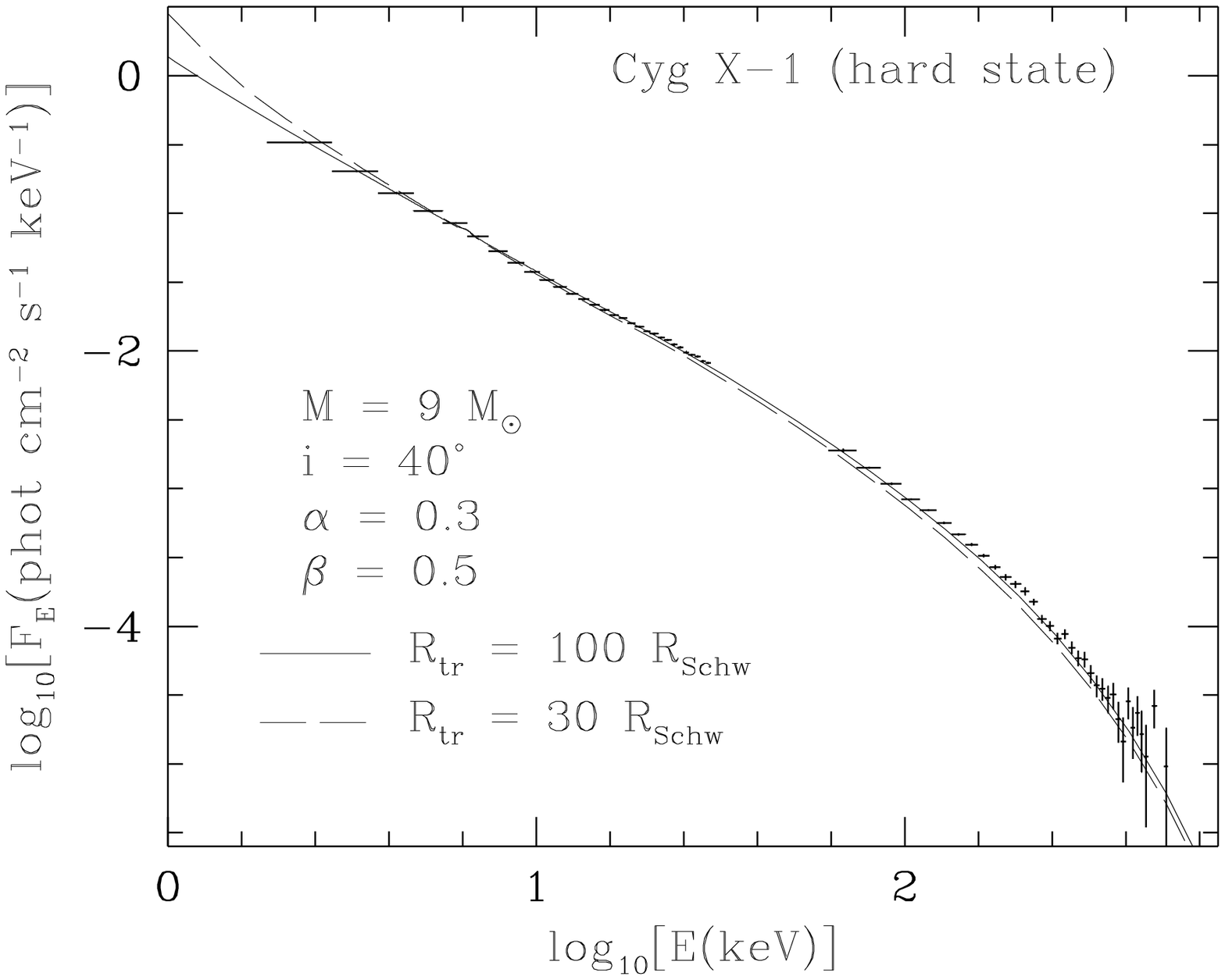} 
\vskip 6.5in 
\caption{Low/hard state spectra of Cyg X-1 observed by Ginga and OSSE
on July 6 1991 (from Gierli\'nski et al. 1997) shown together with two
model spectra corresponding to $\rtr =100$ (solid curve) and $\rtr =
30$ (dashed curve).  Both model spectra are normalized by a factor of
0.37.}
\end{figure}

\begin{figure}
\includegraphics{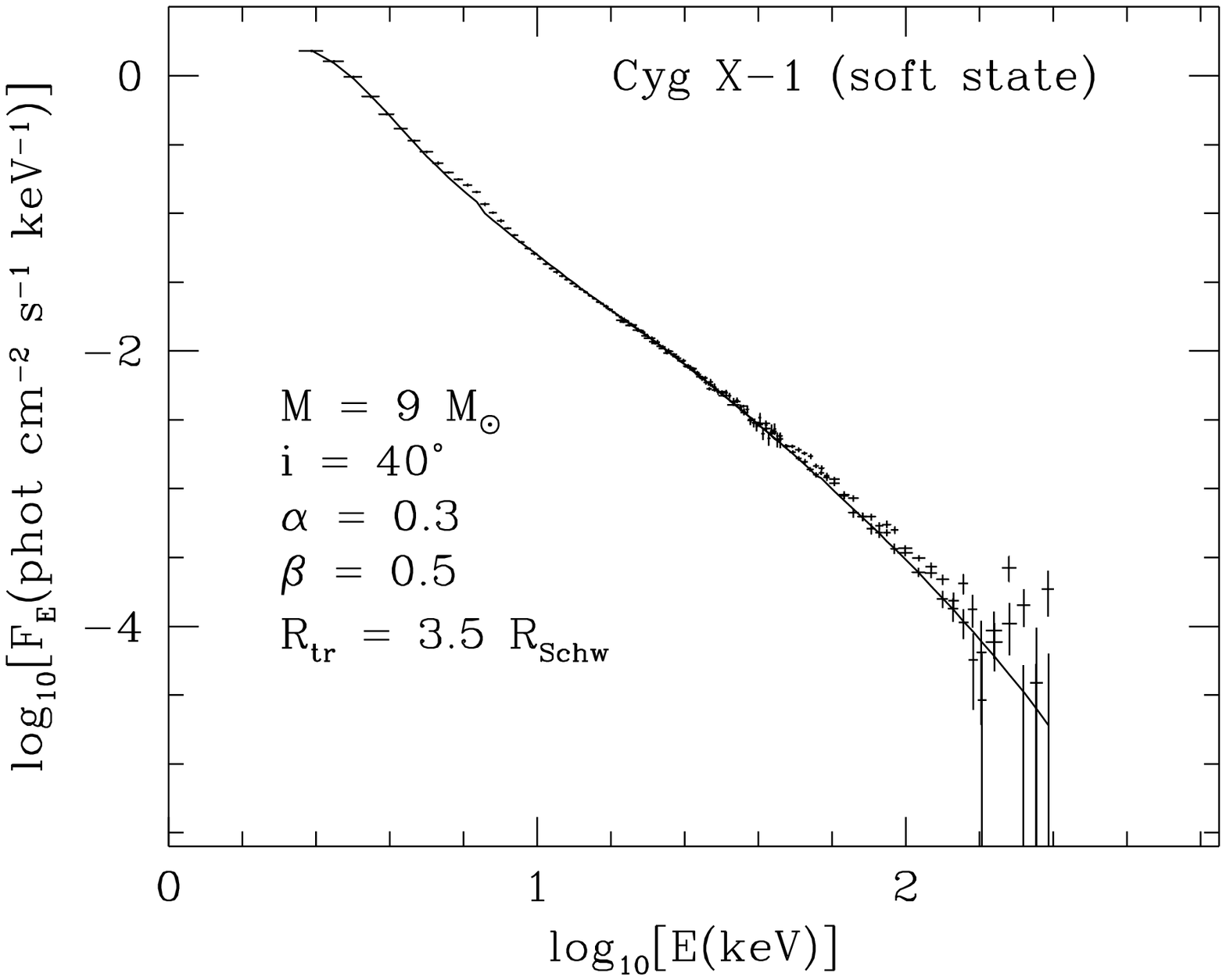} 
\vskip 6.5in 
\caption{Combined PCA/HEXTE high/soft state spectrum of Cyg X-1
observed on 1996 May 22 plotted together with the best fitting model
spectrum corresponding to the intermediate state, normalized by a
factor of 0.48.}
\end{figure}

\begin{figure}
\includegraphics{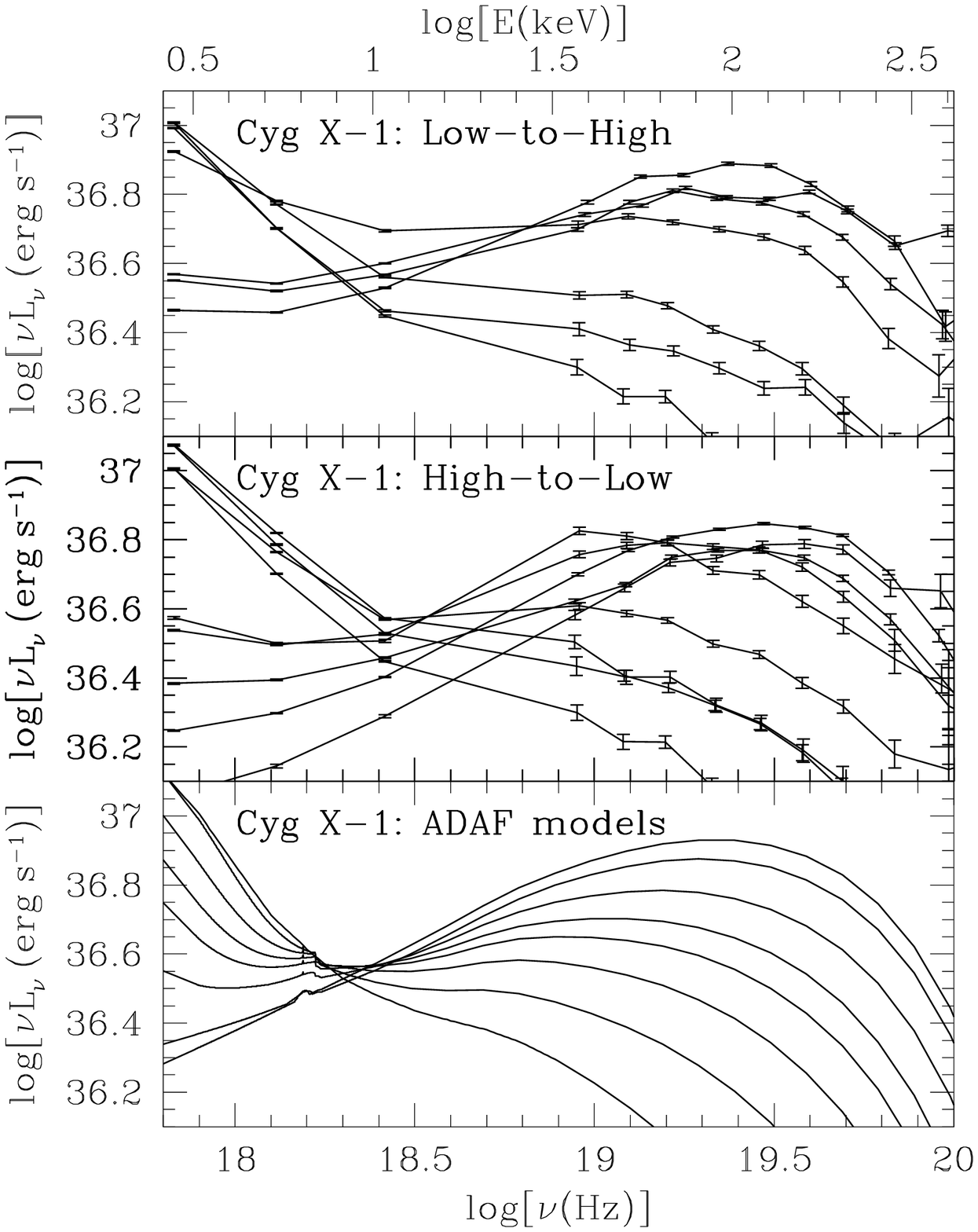} 
\vskip 6.5in 
\caption{Broadband simultaneous RXTE/ASM (1.3-12 keV) and BATSE
(20-600 keV) spectra of Cyg X-1 observed during its 1996 state
transition shown together with the transition sequence predicted by
the EMN model. (a) Spectra observed during the low--high state
transition (TJD 10176-10272).  Note that the lines connecting the data
points are not fits to the data, but are meant simply to guide the
eye.  (b) Spectra observed during the high--low state transition (TJD
10273-10369).  (c) Sequence of intermediate and high state spectra
computed for a model with $m=9$, $i=40^{\circ}$, $\alpha=0.3$,
$\beta=0.5$ and the following values of $\rtr$ (upper to lower spectra
at 100 keV): 100, 30, 10, 6.3, 5, 4, 3.2, 3.  All model spectra have
been normalized by a factor of 0.33.}
\end{figure}

\begin{figure}
\includegraphics{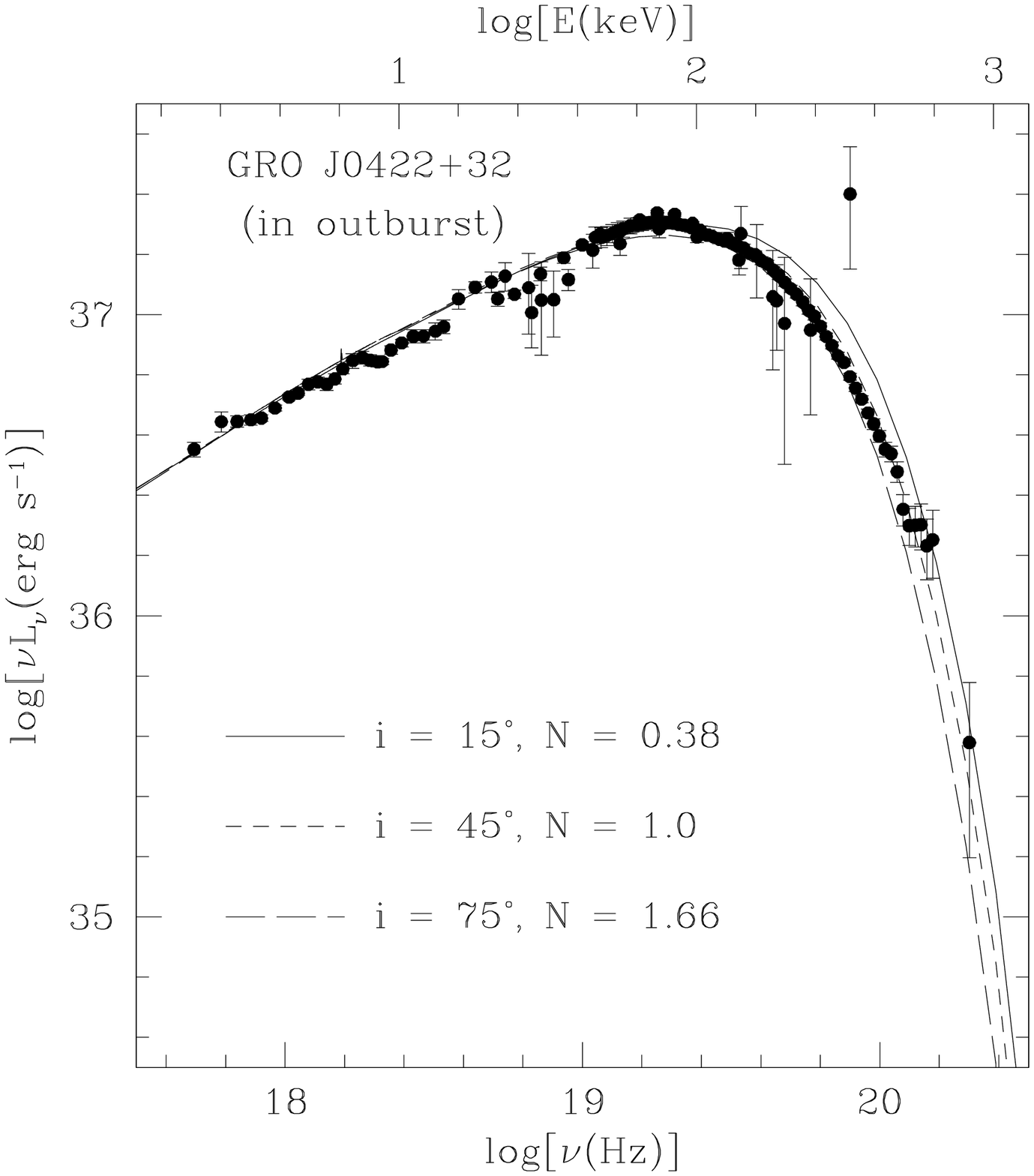} 
\vskip 6.5in 
\caption{The combined TTM (2-20 keV), HEXE (20-200 keV) and OSSE
(50-600 keV) spectrum of GRO J0422+32 observed between 1992 August 29
and September 2, plotted together with the model spectra computed for
$m=9$, $\alpha=0.3$, $\beta=0.5$, $\rtr = 10^4$, $\mdot=0.1$ and
inclination $i$ and normalization factor $N$ as indicated on the
figure.  Though all three spectra fit the data reasonably well, the
one computed with $i=45^{\circ}$ reproduces both the overall
luminosity of the source and the shape of the high energy cut-off.}
\end{figure}

\begin{figure}
\includegraphics{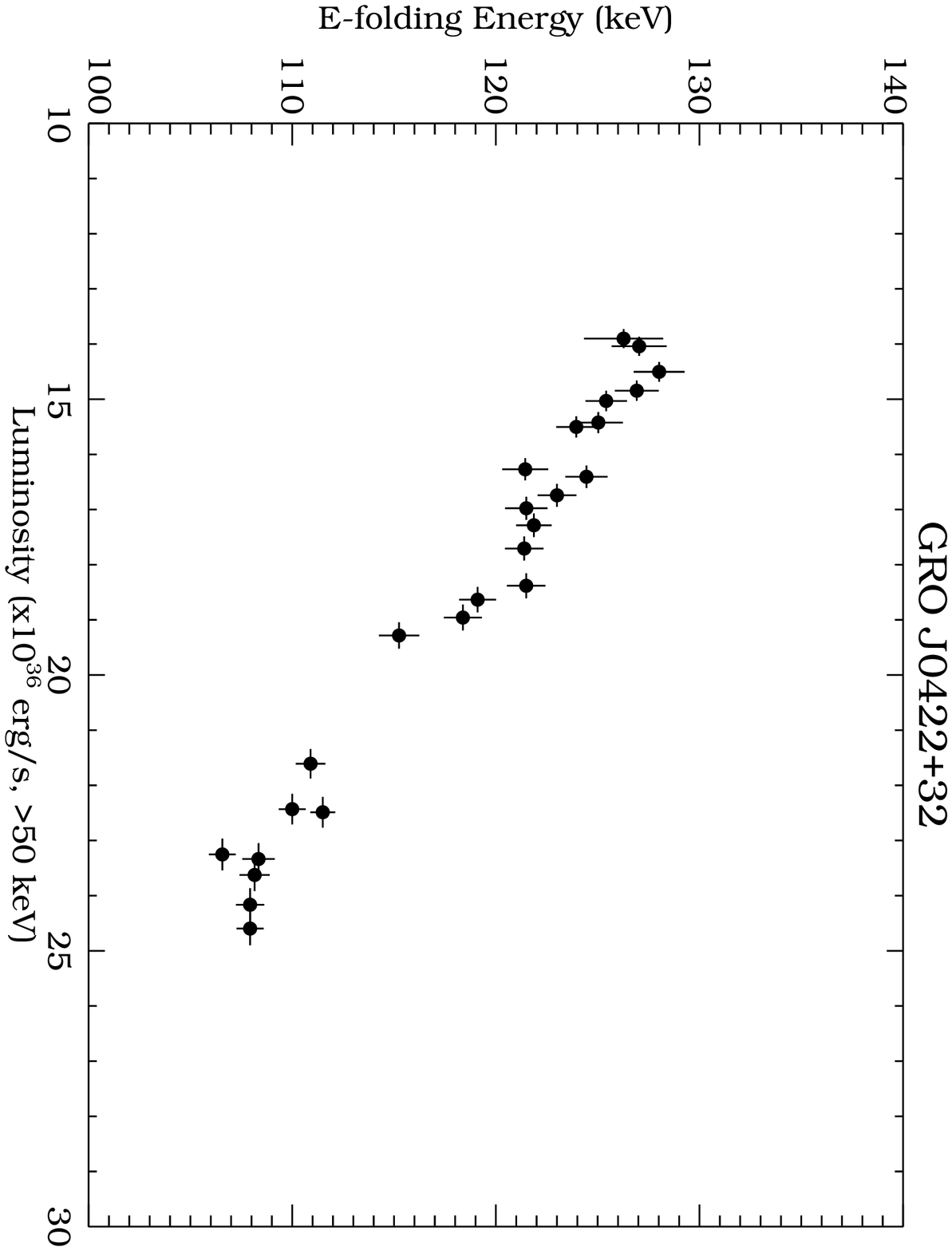} 
\vskip 6.5in 
\caption{The e-folding energy derived from the OSSE spectra of
GRO~J0422+32 plotted against the $\gamma$-ray luminosity (60-1000
keV).  The source was observed between 1992 August 11 and Sept 17,
spanning the interval from the peak of the outburst to approximately
half maximum intensity.  During this time the $\gamma$-ray flux from
GRO J0422+32 decreased monotonically.  Note the clear anticorrelation
between the e-folding energy and high energy flux, indicating that the
source made a transition from the low state toward the quiescent state
(see Figure 2).}
\end{figure}

\begin{figure}
\includegraphics{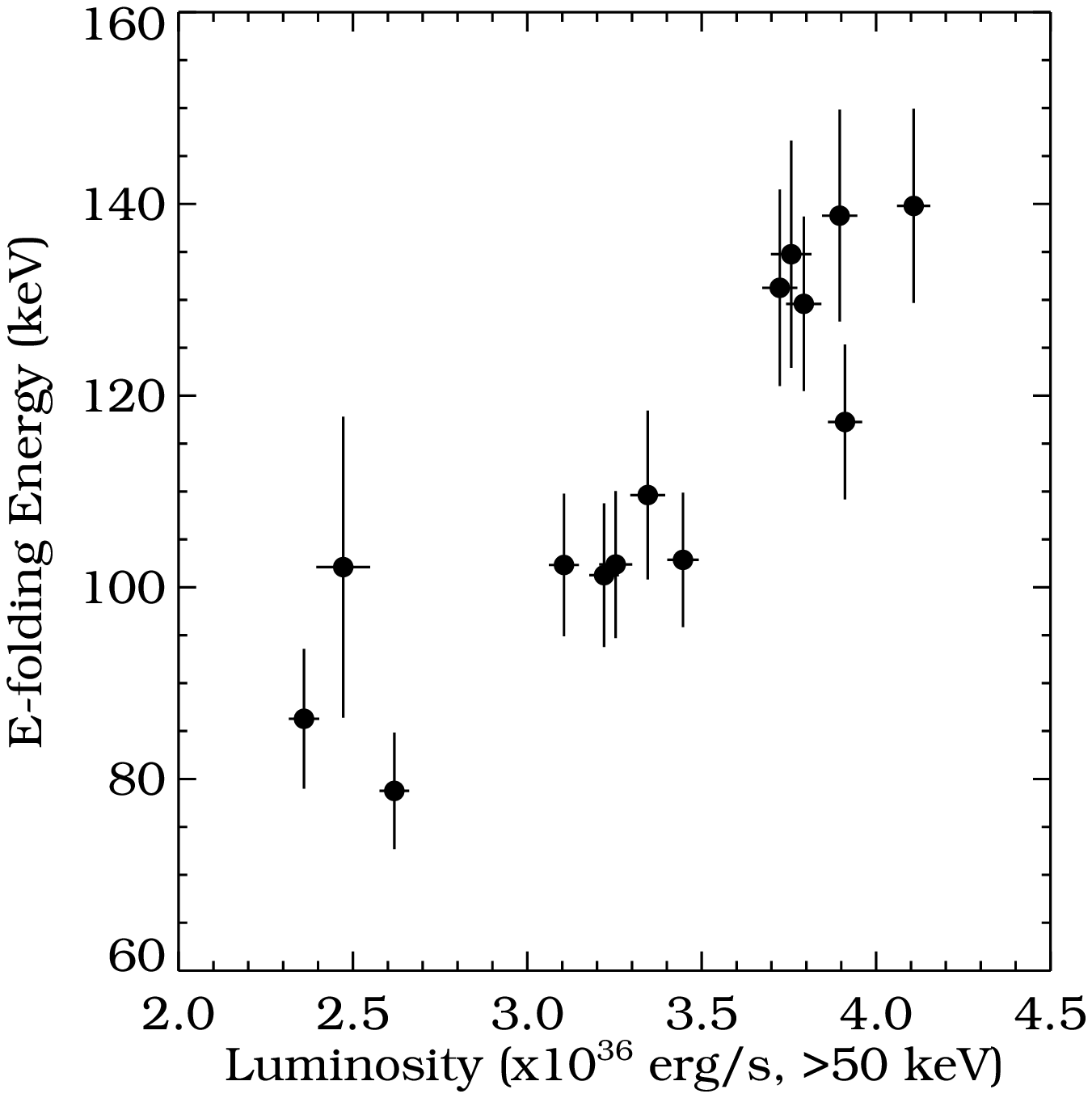} 
\vskip 6.5in 
\caption{The e-folding energy derived from the OSSE spectra of
GRO~J1719$-$24 plotted against the $\gamma$-ray luminosity (60-1000 keV).
The observations were made near the peak of the second smaller flare
(1995 Feb 1-14), during the near monotonic decrease of the
$\gamma$-ray flux.  As opposed to GRO~J0422+32, this source shows an
almost linear correlation between the e-folding energy and high energy
flux, indicating that the source made a transition from the
intermediate state to the high state (see Figure 2).}
\end{figure}

\end{document}